\documentclass[letterpaper,12pt]{article}
\usepackage{graphicx}
\usepackage[hang]{subfigure}
\usepackage{cite}

\def\theequation{\arabic{section}.\arabic{equation}}
\def\theequation{\arabic{section}.\arabic{equation}}
\def\simlt{\stackrel{<}{\scriptscriptstyle\sim}}
\def\simgt{\stackrel{>}{\scriptscriptstyle\sim}}
\def\beq{\begin{equation}}
\def\eeq{\end{equation}}
\def\bea{\begin{eqnarray}}
\def\eea{\end{eqnarray}}
\begin{document}
\vskip -1.5cm
\begin{flushright}
{\small 
ANL-HEP-PR-02-104,~CERN-TH/2002-299,~EFI-02-50,\\[-0.1cm]
FERMILAB-Pub-02/297-T,~MC-TH-2002-10,~MCTP-02-63} \\[-0.1cm] 
hep-ph/0211467, November 2002
\end{flushright}

\begin{center}
{\Large {\bf Collider Probes of the MSSM Higgs Sector}}\\[0.3cm]
{\Large {\bf with Explicit CP Violation}}\\[0.4cm]
{\large M. Carena$^{\,a}$, J. Ellis$^{\,b}$, S. Mrenna$^{\,a,f}$, A.
Pilaftsis$^{\,a,c}$ and C.E.M. Wagner$^{\,d,e}$}\\[0.2cm]
$^a${\em Fermilab, P.O. Box 500, Batavia IL 60510, U.S.A.}\\
$^b${\em Theory Division, CERN, CH-1211 Geneva 23, Switzerland}\\
$^c${\em Department of Physics and Astronomy, University of Manchester,\\ 
Manchester M13 9PL, United Kingdom}\\
$^d${\em High Energy Physics Division, Argonne National Lab., Argonne
  IL 60439, U.S.A.}\\
$^e${\em Enrico Fermi Institute, University of Chicago, 5640 Ellis Ave.,
Chicago IL 60637, U.S.A.}\\
$^f${\it Michigan Center for Theoretical Physics, Randall
Lab.,\\ University of Michigan, Ann Arbor, MI 48109, U.S.A.}

\end{center}

\vskip0.3cm 
\centerline{\bf ABSTRACT} 

{\small  We  investigate  the  hadron collider  phenomenology  of  the
Minimal  Supersymmetric   Standard  Model  (MSSM)   with  explicit  CP
violation  for Higgs  bosons that  can be  observed in  Standard Model
search  channels: $W/ZH_i(\to b\bar  b)$ at  the Tevatron,  and $gg\to
H_i(\to\gamma\gamma)$,   $t\bar  t  H_i(\to   b\bar  b)$   and  $WW\to
H_i(\to\tau^+\tau^-)$ at the LHC.   Our numerical analysis is based on
a  benchmark scenario  proposed  earlier called  CPX,  which has  been
designed to showcase
the effects of CP violation in  the MSSM, and on
several variant benchmarks.  In most of the CPX parameter space, these
hadron  colliders will  find one  of  the neutral  MSSM Higgs  bosons.
However, there are  small regions of parameter space  in which none of
the neutral  Higgs bosons  can be detected  in the  standard channels at 
the Tevatron and the LHC.
This occurs because the neutral  Higgs boson with the largest coupling
to $W$ and $Z$ bosons  decays predominantly  into
either two  lighter Higgs bosons or  a Higgs boson and  a gauge boson,
whilst the lighter Higgs boson has only small couplings to the $W$ and
$Z$ bosons and the top quark.  For other choices of CP-violating parameters,
all three neutral Higgs bosons can have significant couplings to
$W$ and $Z$ bosons, producing overlapping signatures:  these may or may
not be distinguishable from backgrounds.
The  existence  of  these  regions  of parameters  provides  a  strong
motivation for a detailed experimental simulation of these channels. }

\newpage 

\setcounter{equation}{0}
\section{Introduction}\label{sec1}

Detailed studies of the mechanism of electroweak symmetry breaking top
the agendas of operating particle accelerators, as well as those under
construction or being proposed for  the near future.  One paradigm for
many  of  these studies  is  provided  by  the minimal  supersymmetric
extension   of  the   Standard   Model~(MSSM)~\cite{HPN}.   There   is
increasing interest in the possibility that the MSSM includes explicit
CP violation,  
which may provide  opportunities to probe  CP violating
parameters               through               the               Higgs
sector~\cite{APLB,PW,Demir,CDL,CEPW,IN,CPX,CEPW1,CPpp,CPee,CPmumu,CPgeneric}.

Assuming  universality  of   the  soft  supersymmetry  (SUSY)-breaking
gaugino masses $m_{1/2}$, scalar masses $m_0$ and trilinear parameters
$A_0$ at the Grand Unification Theory (GUT) input scale, the MSSM then
provides  only two  new sources  of CP  violation in  addition  to the
Kobayashi--Maskawa phase  of the Standard  Model (SM).  In  a suitable
convention,  these  may  be  represented  as  the  phases  of  complex
parameters  $m_{1/2}$  and $A_0$.   The  values  of  these phases  are
constrained by the upper limits  on the electric dipole moments (EDMs)
of   the    electron,   neutron~\cite{EDMrecent,APEDM}   and   mercury
atom~\cite{FOPR}  and  other measurements.   The
ongoing  probes of CP  violation at  $B$ factories  can also  test the
consistency   of  the   above   minimal  CP-violating   supersymmetric
scenario~\cite{CS,DP}.

In  the MSSM with explicit CP violation,  the effect  of CP  violation on  the Higgs
sector  enters  beyond  the  Born  approximation~\cite{APLB,PW}.   The
effects~\cite{PW,Demir,CDL,CEPW,IN} induced by
arg$(A_0)$ and arg$(m_{1/2})$ on the MSSM Higgs-boson masses have been
studied   in  some   detail~\cite{CEPW,CEPW1}.    Several  interesting
phenomenological  implications for  the  Higgs bosons  emerge in  this
minimal CP-violating  framework of the MSSM.  The  three neutral Higgs
bosons ($h$  and $H$, which have  scalar couplings to  fermions at the
tree  level,  and  $A$   with  pseudoscalar  tree-level  couplings  to
fermions),  all mix  together in  the presence  of CP  violation.  The
resulting three  physical mass eigenstates have mixed  CP parities. We
denote  these by  $H_{1,2,3}$  in order  of  increasing masses,  i.e.,
$M_{H_1} \le M_{H_2} \le M_{H_3}$.   As a consequence of CP violation,
all  the three  neutral  Higgs  bosons can  now  have  tree--level
couplings to pairs of $W^\pm$  and $Z$ bosons.  The Higgs couplings to
the gauge bosons obey the sum rules
\begin{equation}
  \label{one}
\sum^3_{i=1}\ g^2_{H_{i}W^+W^-}\ =\
g^2_{HW^+W^-}\,,\qquad \sum^3_{i=1}\ g^2_{H_iZZ}\ =\ g^2_{HZZ}\,,
\end{equation}
where $g_{HW^+W^-}$ and  $g_{HZZ}$ are  the Standard Model  couplings.
Additionally, there is   an important complementarity  relation between
the $H_iZZ$ and $H_iH_jZ$ couplings:
\begin{equation}
  \label{two}
  g_{H_iH_j Z}\ =\ \varepsilon_{ijk}\, g_{H_kZZ}\,,
\end{equation}
where  $\varepsilon_{ijk}$ is  the  fully anti-symmetric  Levi--Civita
tensor.   We   have  shown  that   the  phenomenological  consequences
of (\ref{one})  and  (\ref{two})   can  be  dramatic~\cite{CEPW}.   In
particular, (\ref{one}) implies that the $H_1ZZ$ coupling can be 
significantly suppressed, thus
raising the possibility that a  relatively light $H_1$ boson,  with a mass 
even  as low as 60~GeV, might have escaped detection at LEP~2. On the 
other 
hand, (\ref{two}) leaves open the possibility of $H_1 H_2$ production at 
LEP~2, as discussed below. However, this brief discussion
highlights the
fact  that  a  large  part  of  the parameter  space  related  to  the
Higgs sector must be re-explored in the presence of CP 
violation.

In this  paper, we investigate systematically the  physics potential of the
Standard Model Higgs boson searches at 
the Tevatron  collider and the LHC  for observing neutral Higgs  bosons in the
MSSM  with  explicit  CP  violation.  
We focus on Standard Model search channels, since the existence of 
a light Higgs boson with a significant coupling to $W$ and $Z$ bosons
is a prediction of weak--scale supersymmetry.
The present work goes beyond previous work on this subject \cite{CPpp} by 
incorporating
the most important radiative corrections to the Higgs sector and 
performing a numerical analysis based on the most realistic simulations
of Standard Model Higgs boson searches at the Tevatron and LHC.
Our  phenomenological  study  is
performed mainly  within the  context of the  CPX scenario~\cite{CPX},
which was  chosen to showcase the  possible effect of  CP violation in
the Higgs sector of the MSSM.   Given the limits from the LEP collider
and existing simulations  of the capabilities of the  Tevatron and the
LHC, we show  that there are small regions of  parameters in which all
three  neutral Higgs  bosons  escape detection.  In  these regions  of
parameters, the heavier Higgs bosons  with the dominant coupling to
the $W$ and $Z$ boson decay into Higgs boson with a tiny coupling to
the $W$ and $Z$ boson, or into
a lighter Higgs boson and  a gauge boson.  Outside these small regions
of parameters,  our analysis indicates, even in the presence of CP 
violation, the  discovery of at  least one neutral
MSSM Higgs  boson in  a set  of  complementary detection
channels.   This provides  a strong  motivation to
extend  the experimental  simulations  to Higgs  boson decay  channels
containing lighter Higgs boson states.  For other choices of CP-violating
parameters, we find that all three neutral Higgs bosons can have
significant couplings to $W$ and $Z$ bosons, while being closely spaced
in mass.  This situation also requires careful investigation, since 
some signals may be affecting the background estimates for other
signals.

Our   numerical   analysis   is   performed  using   the   code   {\tt
CPHDECAY}~\cite{CPHDECAY}, which is based  on an extension of the code
{\tt  HDECAY}~\cite{HDECAY}   to  calculate  the   Higgs  boson  decay
properties, and  of the code  {\tt cph}~\cite{CEPW, CPH}  to calculate
the  physical  Higgs  spectrum  and  mixing angles.  The  latter  code
includes  the dominant  one- and  two-loop  CP-violating contributions
induced  by   the  supersymmetric  particles.   At   one  loop,  these
contributions  depend  mostly  on  the  phase of  the  trilinear  mass
parameters  of the  third-generation quarks,  $A_{t,b}$,  relative to
that  of the  supersymmetric mass  parameter $\mu^{*}$.   At  the two-loop
level, a  significant dependence on  the relative phase  between $\mu^{*}$
and the  gluino mass parameter $m_{\tilde{g}}$ also  appears.  Electroweak corrections
are  incorporated  at  the  leading-logarithmic  level.   In  a  phase
convention where the $\mu$ parameter is positive, the input parameters
for the  code are the top-quark  mass $m_t$, the  $\mu$ parameter, the
soft      supersymmetry-breaking      masses      $\widetilde{M}^2_Q$,
$\widetilde{M}^2_{t,b}$   of   the   third-generation   squarks,   the
third-generation  soft  trilinear  couplings $A_{t,b}$  together  with
their  respective phases,  the gaugino  masses  and the  phase of  the
gluino  mass  parameter   $m_{\tilde{g}}$.   Apart  from  its  present
application to  hadron colliders, {\tt  CPHDECAY} may also
serve as a useful tool for analogous studies at future
linear $e^+e^-$
colliders~\cite{CPee} or a muon collider~\cite{CPmumu}, or for
generic Higgs boson studies~\cite{CPgeneric}.

Section~\ref{sec2}  contains a  complete discussion  of  the effective
couplings of the neutral and charged Higgs bosons to quarks and of the
Higgs self-couplings;  full expressions are given in  Appendix A.  The
one-loop corrected  Higgs couplings  are essential, as  they determine
the  branching  ratios  of  the Higgs  bosons in the presence of CP 
violation.
Section~\ref{sec3}  contains a discussion  of the  LEP results  in the
search  for  neutral  Higgs  bosons  in  the  MSSM  with  explicit  CP
violation.   Section~\ref{sec4} contains a  numerical analysis  of the
different  collider  detection  channels,  showing  graphically  their
complementary  properties  for  the  detection of Standard Model--like  
Higgs bosons in the MSSM with explicit CP violation.    
We  restrict   ourselves  to   the   existing  experimental
simulations,  identifying regions  of  parameters at  small values  of
$\tan\beta$ and of the charged Higgs boson mass, for which none of the
neutral Higgs bosons  can be detected at the Tevatron  or the LHC via
the standard search channels. Our
conclusions are  presented in Section~\ref{sec5}.  Appendix A contains
detailed  formulae  for  couplings   and  Appendix  B  summarizes  the
information from detector  simulations 
on which we base our analysis~\cite{Carena:2000yx,LHC2,Drollinger:2001ym,Lassila-Perini}.

\setcounter{equation}{0}
\section{Effective Higgs-Boson Couplings}\label{sec2} 

In order to study the collider phenomenology of the Higgs bosons in
the MSSM with explicit CP violation, we must first calculate the
couplings of the Higgs bosons to Standard Model and MSSM particles.
In this section, we present explicitly those couplings which are the
most important inputs to our numerical calculations.  In particular,
we review the effective couplings of the neutral Higgs bosons to
quarks, and we present the effective Higgs self-couplings.  We also
give the corresponding effective couplings of the charged Higgs boson
to the up and down quarks, even though they are not needed for the
present analysis.

\subsection{Effective Higgs-boson couplings to quarks}

Following our previous conventions   throughout this paper \cite{CEPW}, we
define   the physical  scalar  components   of the   Higgs superfields
$\widehat{H}_1$   and   $\widehat{H}_2$  as   $\widetilde{\Phi}_1    =
i\tau_2\,\Phi^*_1 =  (\phi^{0*}_1,\  -\,\phi^-_1   )$ and  $\Phi_2   =
(\phi^+_2,\  \phi^0_2 )$,  respectively, where  $\tau_2$  is the usual
Pauli  matrix.  The weak Higgs eigenstates  are  related to their mass
eigenstates   $H^+$    and   $H_{1,2,3}$    through the     orthogonal
transformations:
\begin{eqnarray}
  \label{Phiplus}
\phi^+_1 &=& \cos\beta\; G^+\:  -\: \sin\beta \; H^+\,,\qquad
\phi^+_2 \ =\ \sin\beta\; G^+\:  +\: \cos\beta \; H^+\,,\\
  \label{Phi0}
\phi_1^0 &=& {\textstyle \frac{1}{\sqrt{2}}}\, \Big[\, v_1\ +\ 
O_{1i}\, H_i\ +\ i\, 
\Big( \cos\beta ~G^0\: -\: \sin\beta ~O_{3i}\, H_i\,\Big)\, \Big]\,,
  \nonumber \\
\phi_2^0 &=& {\textstyle \frac{1}{\sqrt{2}}}\, \Big[\, v_2\ +\ 
O_{2i}\, H_i\ +\ i\, 
\Big( \sin\beta ~G^0\: +\: \cos\beta ~O_{3i}\, H_i\,\Big)\, \Big] \,,
\end{eqnarray}
where  $G^+$ and  $G^0$  are the would-be  Goldstone bosons associated
with  the  $W^+$ and  $Z$ bosons,  respectively,  and  $O$ is a 3-by-3
orthogonal  matrix that  describes  the  mixing  of the neutral  Higgs
states in the presence of CP violation~\cite{PW}.

Our starting point is the effective Lagrangian for the neutral Higgs-boson
couplings to the $u$- and $d$-type quarks.  We neglect quark-mixing
effects, as their direct relevance to Higgs searches is not
important.\footnote{For a recent discussion of CP-violating Higgs-mediated
flavor-changing neutral-current effects on low-energy observables,
see Ref.~\cite{DP}.} Moreover, we assume that there is negligible mixing
between the different generations of squarks, and we include
weak-interaction effects only at the leading-logarithmic level.

Under   these  assumptions,  the
effective Lagrangian reads:
\begin{eqnarray}
  \label{eff0}
-\,{\cal L}^0_{\rm eff} &=& \Big[\, (h_d\: +\: \delta h_d)\,
\phi^{0*}_1\: +\: \Delta h_d\, \phi^{0*}_2\, \Big]\,
\bar{d}_R\,d_L\nonumber\\
&& +\: \Big[\, \Delta h_u\, \phi^{0}_1\: +\: 
(h_u\: +\: \delta h_u)\, \phi^{0}_2\,\Big]\, \bar{u}_R\,u_L\ +\
{\rm h.c.},
\end{eqnarray}
where  $\delta  h_{d,u}/h_{d,u}$   and  $\Delta h_{d,u}/h_{d,u}$   are
threshold radiative effects given by
\begin{eqnarray}
  \label{dhb}
\frac{\delta h_d}{h_d} &=& -\frac{2 \alpha_s}{3 \pi}\, m_{\tilde{g}}^* A_d\,
I(m^2_{\tilde{d}_1},m^2_{\tilde{d}_2},|m_{\tilde{g}}|^2)\: -\: 
\frac{|h_u|^2}{16 \pi^2}\, |\mu|^2\,
I(m^2_{\tilde{u}_1},m^2_{\tilde{u}_2},|\mu|^2)\, ,\\
  \label{Dhb}
\frac{\Delta h_d}{h_d} &=& 
\frac{2 \alpha_s}{3\pi}\, m^*_{\tilde{g}}\, \mu^* I(m_{\tilde{d}_1}^2,
m_{\tilde{d}_2}^2,|m_{\tilde{g}}|^2)\ +\ \frac{|h_u|^2}{16\pi^2}\, 
A^*_u \mu^*\, I(m_{\tilde{u}_1}^2,m_{\tilde{u}_2}^2,|\mu|^2)\, ,\\
  \label{Dht}
\frac{\Delta h_u}{h_u} &=& 
\frac{2 \alpha_s}{3\pi}\, m^*_{\tilde{g}} \mu^*\, I(m_{\tilde{u}_1}^2,
m_{\tilde{u}_2}^2,|m_{\tilde{g}}|^2)\ +\ \frac{|h_d|^2}{16\pi^2}\, 
A_d^* \mu^*\, I(m_{\tilde{d}_1}^2,m_{\tilde{d}_2}^2,|\mu|^2)\, ,\\
  \label{dht}
\frac{\delta h_u}{h_u} &=& 
-\frac{2 \alpha_s}{3\pi}\, m^*_{\tilde{g}} A_u\, I(m_{\tilde{u}_1}^2,
m_{\tilde{u}_2}^2,|m_{\tilde{g}}|^2)\ 
-\ \frac{|h_d|^2}{16\pi^2}\, |\mu|^2\,
I(m_{\tilde{d}_1}^2,m_{\tilde{d}_2}^2,|\mu|^2)\, .
\end{eqnarray}
We would like to emphasize that no other work includes the $\Delta h_b$
corrections, and that these have important phenomenological implications, 
as we discuss below.

In~(\ref{dhb})--(\ref{dht}),  $\alpha_s   =   g^2_s/(4\pi)$  is    the
SU(3)$_c$ fine-structure constant, $I(a,b,c)$ is the function
\begin{equation}
I(a,b,c)\ \equiv \ \frac{ a b \ln (a/b) + b c \ln (b/c) + a c \ln (c/a)}
{(a-b)(b-c)(a-c)}\ ,
\end{equation}
and  $m_{\tilde{q}_1,  \tilde{q}_2}$  (with  $q=u,d$)  are the  squark
masses, with
\begin{eqnarray}
  \label{mq12}         
m^2_{\tilde{q}_1\,    (\tilde{q}_2)} \!\!&=&\!\! 
  \frac{1}{2}\, \bigg\{\,
\widetilde{M}^2_Q\: +\: \widetilde{M}^2_q\: +\: 2\,m^2_q\: +\: 
T^q_z\,\cos 2\beta M^2_Z \\
&&\!\! +(-)\, \sqrt{\Big[\, \widetilde{M}^2_Q - \widetilde{M}^2_q +
\cos 2\beta M^2_Z\, (T^q_z - 2 Q_q \sin^2\theta_w)\, \Big]^2\: +\:
4\,m^2_q\,|A_q - R_q \mu^* |^2\, }\ \bigg\}\, ,\nonumber
\end{eqnarray}
and $Q_u\,  (Q_d)  = 2/3\, (-1/3)$,  $T^u_z  = - T^d_z  = 1/2$, $R_u\,
(R_d)  =  \cot\beta \    (\tan\beta)$,   and $\sin^2\theta_w =  1    -
M^2_W/M^2_Z$. In~(\ref{eff0}), the  complex Yukawa couplings $h_d$ and
$h_u$ are defined in a way such that the $d$- and $u$-quark masses are
real and positive after the inclusion of radiative corrections, namely
\begin{eqnarray}
  \label{hdu}
h_d &=& \frac{m_d}{v \cos\beta} \; \frac{1}{1\, +\, (\delta h_d/h_d)\,
+\, (\Delta h_d/h_d) \tan\beta}\ ,\\
h_u &=& \frac{m_u}{v \sin\beta} \; \frac{1}{1\, +\, (\delta h_u/h_u) 
\, +\, (\Delta h_u/h_u) \cot\beta}\ .
\end{eqnarray}
Substituting (\ref{Phi0})    and  (\ref{hdu}) into~(\ref{eff0}),   the
effective Lagrangian for  the  $H_i\bar{q}q$ couplings can be  written
in the general form:
\begin{equation}
  \label{Hqq}
{\cal L}_{H_i\bar{q}q}\ =\ - \sum_{q=u,d}\,\frac{g_w m_q}{2 M_W}\,
\sum_{i=1}^3\, H_i\, \bar{q}\,\Big( g^S_{H_i\bar{q}q}\, +\,
ig^P_{H_i\bar{q}q}\gamma_5 \Big)\, q\ ,
\end{equation}
with~\cite{CEPW,APEDM}
\begin{eqnarray}
  \label{gSHbb}
g^S_{H_i\bar{d}d} & =& {\rm Re}\, \bigg(\,
\frac{1}{1\, +\, \kappa_d\,\tan\beta}\,\bigg)\,
\frac{O_{1i}}{\cos\beta}
\ +\ {\rm Re}\, \bigg(\, \frac{\kappa_d}{1\, +\, 
\kappa_d\, \tan\beta}\,\bigg)\ 
\frac{O_{2i}}{\cos\beta}\nonumber\\
&& +\: {\rm Im}\, \bigg[\,
\frac{ \kappa_d\, (\tan^2\beta\, +\, 1)}{1\, +\, 
\kappa_d\, \tan\beta}\,\bigg]\
O_{3i}\, , \\[0.35cm]
  \label{gPHbb}
g^P_{H_i\bar{d}d} & =& -\, {\rm Re}\, \bigg(\, 
\frac{ \tan\beta\, -\, \kappa_d}{1\, +\, \kappa_d \tan\beta}\,\bigg)\, O_{3i}
\ +\ {\rm Im}\, \bigg(\, \frac{\kappa_d\,\tan\beta}{1\, +\, 
\kappa_d\, \tan\beta}\,\bigg)\ 
\frac{O_{1i}}{\cos\beta}\nonumber\\
&&-\: {\rm Im}\, \bigg(\,
\frac{\kappa_d}{1\, +\, \kappa_d\, \tan\beta}\,\bigg)\
\frac{O_{2i}}{\cos\beta}\ , \\[0.35cm]
  \label{gSHtt}
g^S_{H_i\bar{u}u} & =& {\rm Re}\, \bigg(\,
\frac{1}{1\, +\, \kappa_u\,\cot\beta}\,\bigg)\, \frac{O_{2i}}{\sin\beta}
\ +\ {\rm Re}\, \bigg(\, \frac{\kappa_u}{1\, +\, 
\kappa_u\, \cot\beta}\,\bigg)\ 
\frac{O_{1i}}{\sin\beta}\nonumber\\
&& +\: {\rm Im}\, \bigg[\,
\frac{\kappa_u\, (\cot^2\beta\, +\, 1)}{1\, +\, 
\kappa_u\, \cot\beta}\,\bigg]\
O_{3i}\, , \\[0.35cm]
  \label{gPHtt}
g^P_{H_i\bar{u}u} & =& -\, {\rm Re}\, \bigg(\,
\frac{\cot\beta\, -\, \kappa_u}{1\, +\, 
\kappa_u\,\cot\beta}\,\bigg)\, O_{3i}\ +\  
{\rm Im}\, \bigg(\, \frac{\kappa_u\,\cot\beta}{1\, +\, 
\kappa_u\, \cot\beta}\,\bigg)\ 
\frac{O_{2i}}{\sin\beta}\nonumber\\
&& -\: {\rm Im}\, \bigg(\,
\frac{\kappa_u}{1\, +\, \kappa_u\, \cot\beta}\,\bigg)\
\frac{O_{1i}}{\sin\beta}\ .
\end{eqnarray}
In the above, we have used the abbreviation:  $\kappa_q = (\Delta h_q/
h_q)/[1 + (\delta h_q/h_q)]$, for $q=u,d$, and the  fact that $u$- and
$d$-quark masses are real and positive, so that ${\rm Im}\, m_d  \propto  
{\rm
Im}\, [ h_d + (\delta  h_d) + (\Delta h_d)  \tan\beta ] = 0$ and ${\rm
Im}\,  m_u \propto  {\rm  Im}\, [ h_u  +  (\delta h_u) +  (\Delta h_u)
\cot\beta ] = 0$.

Though our phenomenological analysis is restricted to the
neutral Higgs-boson sector, we consider for completeness
the charged--Higgs--quark sector.   The
effective Lagrangian describing the interactions  of the charged--Higgs
fields $\phi_{1,2}^+$ with $u$ and $d$ quarks is given by
\begin{eqnarray}
  \label{effplus}
-\,{\cal L}^\pm_{\rm eff} &=& \Big[\,
(h_d\: +\: \bar{\delta} h_d)\, \phi^{-}_1\: 
+\: \bar{\Delta} h_d\, \phi^{-}_2\,\Big]\,
\bar{d}_R\,u_L\nonumber\\
&& -\: \Big[\, \bar{\Delta} h_u\, \phi^{+}_1\: +\: 
(h_u\: +\: \bar{\delta} h_u)\, \phi^{+}_2\, \Big]\, \bar{u}_R\,d_L\ +\
{\rm h.c.}
\end{eqnarray}
The  quantities  $\bar{\delta}   h_{u,d}$  and  $\bar{\Delta} h_{u,d}$
contain the  respective threshold radiative   effects that affect  the
charged--Higgs couplings to quarks. Though  they are related to the
$\delta  h_q$ and  $\Delta h_q$ given above for the neutral Higgs
sector, they differ by SU(2)$_L$-breaking terms.

To facilitate our presentation of  the analytic forms of $\bar{\delta}
h_q$  and $\bar{\Delta} h_q$, we first  set up our conventions for the
squark-mixing parameters   and define some  auxiliary functions. Thus,
the   left-    and  right-handed  squark    fields  $\tilde{q}_L$  and
$\tilde{q}_R$ are  related to  the physical   fields $\tilde{q}_{1,2}$
through the transformations:
\begin{eqnarray}
\tilde{q}_L & = & c_q\, \tilde{q}_1\: +\: s_q\, \tilde{q}_2
\nonumber\\
\tilde{q}_R & = & -\, e^{i\delta_q}\, s_q\, \tilde{q}_1\: +\:
e^{i\delta_q}\, c_q\, \tilde{q}_2
\end{eqnarray}
where  $\delta_q = {\rm arg}\, (  A_q - R_q  \mu^*  )$ and $c_q = \cos
\theta_{\tilde{q}}$   and   $s_q   =   \sin   \theta_{\tilde{q}}$  are
squark-mixing angles.  Analytic expressions for $c_q$ and $s_q$ may be
found in Appendix~A of~\cite{CEPW1}.
We introduce   the following short-hand   form  for  the squark-loop
integrals:
\begin{equation}
  \label{Iijk}
I(i,j,m)\  =\ I(m^2_{\tilde{u}_i},m^2_{\tilde{d}_j},|m|^2)\, ,
\end{equation}
where the indices $i$ and $j$ refer to the third-generation
up- and down-squark mass eigenstates.

With the aid  of~(\ref{Iijk}), the  following two auxiliary  functions
can be defined:
\begin{eqnarray}
  \label{I1}
I_1(m) &=& c_u^2 s_d^2 I(1,1,m)\: +\: s_u^2 c_d^2 I(2,2,m)\:
+\: c_u^2 c_d^2 I(1,2,m)\: +\: s_u^2 s_d^2 I(2,1,m)\, ,\qquad\\
  \label{I2}
I_2(m) &=& s_u^2 c_d^2 I(1,1,m)\: +\: c_u^2 s_d^2 I(2,2,m) 
\: +\: s_u^2 s_d^2 I(1,2,m)\: +\: c_u^2 c_d^2 I(2,1,m)\, .
\end{eqnarray}
Again, in the limit of neglecting mixing of generations, the threshold
radiative effects related to the charged Higgs sector can be expressed
in terms of~(\ref{I1}) and~(\ref{I2}) as follows:
\begin{eqnarray}
  \label{dhbp}
\frac{\bar{\delta} h_d}{h_d} & =& -\, \frac{2 \alpha_s}{3 \pi}\, 
m_{\tilde{g}}^* A_d\, I_1(m_{\tilde{g}})\  -\ \frac{|h_u|^2}{16 \pi^2}\,
|\mu|^2\, I_2(\mu)\, ,\\
  \label{Dhbp}
\frac{\bar{\Delta} h_d}{h_d} &=& \frac{2 \alpha_s}{3 \pi}\,
m_{\tilde{g}}^* \mu^*\, I_1(m_{\tilde{g}})\  +\ \frac{|h_u|^2}{16 \pi^2}\,
\mu^* A_u^*\, I_2(\mu)\, ,\\
  \label{dhtp}
\frac{\bar{\delta} h_u}{h_u} &=& - \frac{2 \alpha_s}{3 \pi}\, 
m_{\tilde{g}}^* A_u\, I_2(m_{\tilde{g}})\  -\ \frac{|h_d|^2}{16 \pi^2}\,
|\mu|^2\, I_1(\mu)\, ,\\
  \label{Dhtp}
\frac{\bar{\Delta} h_u}{h_u} &=& \frac{2 \alpha_s}{3 \pi}\, 
m_{\tilde{g}}^* \mu^*\, I_2(m_{\tilde{g}})\  +\ \frac{|h_d|^2}{16 \pi^2}\,
\mu^* A_b^*\, I_1(\mu)\, .
\end{eqnarray}
Then,  the effective Lagrangian for  the charged Higgs-boson couplings
to $u$- and $d$-type quarks is given by
\begin{equation}
  \label{Hud}
{\cal L}_{H^\pm ud}\ =\ \frac{g_w}{2 M_W}\, H^+\, 
\bar{u}\,\Big(\, m_u\, g^L_{H^+\bar{u}d}\, P_{-}\ +\ m_d\,
g^R_{H^+\bar{u}d}\, P_{+}\, \Big)\, d\ +\ {\rm h.c.} ,
\end{equation}
where $P_{\mp} = [1 \mp \gamma_5 ]/2$, and 
\begin{eqnarray}
  \label{gLHud}
g^L_{H^+\bar{u}d} & =& \frac{\cot\beta\, (1\, +\, \rho_u )\:
-\: \bar{\kappa}_u}{1\: +\: \kappa_u\,\cot\beta} \ , \\[0.35cm]
  \label{gRHud}
g^R_{H^+\bar{u}d} & =& \frac{\tan\beta\, (1\, +\, \rho^*_d )\:
-\: \bar{\kappa}^*_d}{1\: +\: \kappa_d^*\,\tan\beta} \ .
\end{eqnarray}
The quantities $\kappa_{u,d}$   are given  after~(\ref{gPHtt}),  while
$\bar{\kappa}_{u,d}$   and    $\rho_{u,d}$    in~(\ref{gLHud})
and~(\ref{gRHud}) are defined as follows:
\begin{equation}
\bar{\kappa}_{q}\ =\ \frac{(\bar{\Delta} h_q/h_q )}{1 + (\delta
h_q/h_q)}\ ,\qquad \rho_q\ =\ \frac{(\bar{\delta} h_q/h_q)\:
-\: (\delta h_q/h_q)}{1 + (\delta h_q/h_q)}\ ,
\end{equation}
with $q=u,d$, respectively.

It is apparent that the effective couplings of the neutral and charged
Higgs bosons to quarks can be  written in a form that depends entirely
on    the   parameters   $\kappa_{u,d}$,    $\bar{\kappa}_{u,d}$   and
$\rho_{u,d}$.  It  is technically interesting  to note that  all these
quantities  are renormalization-scale independent  up to  the one-loop
order in our resummation approach. More explicitly, $\kappa_{u,d}$ and
$\bar{\kappa}_{u,d}$ are proportional to the non-holomorphic radiative
corrections    $(\Delta     h_{u,d}/h_{u,d})$    and    $(\bar{\Delta}
h_{u,d}/h_{u,d})$,     respectively,      which     are     manifestly
scale-independent.   The  quantities   $\rho_{u,d}$  are  measures  of
SU(2)$_L$ breaking in the up- and down-Yukawa sectors.  The parameters
$\rho_{u,d}$   are  also   renormalization-scale  independent,   as  a
consequence of  the SU(2)$_L$ gauge invariance of  the original theory
before  spontaneous  symmetry  breaking.   The scale  independence  of
$\rho_{u,d}$  can  be also  understood  by  simply  noticing that  the
ultra-violet  infinities, e.g.,  of $\delta  h_{d}$  and $\bar{\delta}
h_{d}$ are equal,  as they emanate from the  charged and neutral Higgs
components of the same gauge-invariant operator, which is $\widehat{Q}
\widehat{H}_1 \widehat{D}$  in this  case.  The scale  independence of
these  quantities has  some analogies  with that  of  Veltman's $\rho$
parameter  that characterizes  the isospin  breaking in  the  SM gauge
sector through the difference between the $WW$ and $ZZ$ self-energies,
that we do not develop further here.

Finally, we  should bear  in mind that  the validity of  the effective
neutral  and charged  Higgs-boson couplings  to quarks  depend  on two
kinematic  conditions: (i)~the  soft  supersym- metry-breaking  masses
should be much larger than the electroweak scale and (ii)~the external
momenta of  the quarks and Higgs  bosons, e.g., $p_{q,H}$,  have to be
sufficiently smaller  than the soft  supersymmetry-breaking mass scale
$M_{\rm  SUSY}$, so  that they  can  be neglected  when expanding  the
vertex functions in powers of $p^2_{q,H}/M^2_{\rm SUSY}$.

\subsection{Effective Higgs-boson self-couplings}

To determine the effective Higgs self-couplings, one needs to know the
analytic forms of  both the proper vertex  and self-energy graphs.  So
far, there is  complete information only  for the latter contributions
in  the  MSSM  with  explicit CP violation~\cite{CEPW,CEPW1},  while  the
former effects have been computed in~\cite{PW}, in an expansion of the
effective potential up to operators of dimension 4.

Here,  we  combine the  above  two  pieces  of information  to  obtain
approximate  analytic expressions for  the trilinear  and quadrilinear
Higgs-boson  self-couplings.\footnote{A  similar  procedure  has  been
followed in Ref.~\cite{CL}.}  Our analytic expressions are not limited
to  the MSSM  case, but  can be  applied equally  well to  the general
Two-Higgs-Doublet Model (2HDM) with explicit CP violation.

To start with, we first write down the effective Lagrangian containing
all operators of dimension 4:
\begin{eqnarray}
  \label{Leff4}
{\cal L}^{\rm 4d}_{\rm eff} & = & \lambda_1\, (\Phi^\dagger_1
\Phi_1)^2\: +\: \lambda_2\, (\Phi^\dagger_2 \Phi_2)^2\: +\:
\lambda_3\, (\Phi^\dagger_1 \Phi_1)\, (\Phi^\dagger_2
\Phi_2)\: +\: \lambda_4\, (\Phi^\dagger_1\Phi_2)\, 
(\Phi^\dagger_2\Phi_1)\nonumber\\
&&+\: \lambda_5\, (\Phi^\dagger_1 \Phi_2)^2 \: +\:
\lambda^*_5\, (\Phi^\dagger_2 \Phi_1)^2 \: +\: 
\lambda_6\, (\Phi^\dagger_1\Phi_1)\, (\Phi^\dagger_1\Phi_2 )\: +\:
\lambda^*_6\, (\Phi^\dagger_1\Phi_1)\, (\Phi^\dagger_2\Phi_1)\nonumber\\
&&+\: \lambda_7\, (\Phi^\dagger_2\Phi_2)\, (\Phi^\dagger_1\Phi_2 )\: +\:
\lambda^*_7\, (\Phi^\dagger_2\Phi_2)\, (\Phi^\dagger_2\Phi_1 )\, ,
\end{eqnarray}
where $\Phi^T_{1,2}\ =\  (\phi^+_{1,2}\,,\ \phi^0_{1,2} )$ are the two
Higgs doublets, with  their individual charged and  neutral components
defined   in~(\ref{Phiplus})  and (\ref{Phi0}).  After (\ref{Phiplus})
and (\ref{Phi0}) have  been inserted into (\ref{Leff4}), the effective
trilinear and  quartic Higgs self-couplings may  be cast into the form
(in the unitary gauge):
\begin{eqnarray}
  \label{H3}
{\cal L}^{3H}_{\rm eff} & =&  v\, \Big(\, \Gamma^{3H}_{ijk}\, H_i H_j
H_k\: +\: \Gamma^{HH^+H^-}_i\, H_i H^+ H^- \,\Big)\,,\\
  \label{H4}
{\cal L}^{4H}_{\rm eff} & =& \Gamma^{4H}_{ijkl}\, H_i H_j H_k H_l\
+\  \Gamma^{2H H^+H^-}_{ij}\, H_i H_j H^+ H^-\ +\
\Gamma^{4H^+}\, (H^+ H^-)^2\,,\qquad
\end{eqnarray}
where $v = \sqrt{v^2_1 + v^2_2} \simeq 246$~GeV is the SM vacuum
expectation value. The couplings $\Gamma^{3H}_{ijk}$ and
$\Gamma^{4H^+}$ are given by 
\begin{eqnarray}
  \label{coupls}
\Gamma^{3H}_{ijk} \!\!\!&=&\!\!\! \sum\limits_{l\le m \le n = 1,2,3}\
O_{li}\, O_{mj}\, O_{nk}\ g^{3H}_{lmn}\,,\qquad
\Gamma^{HH^+H^-}_i =\  \sum\limits_{l= 1,2,3}\ 
O_{li}\ g^{HH^+H^-}_{l}\,,\\[3mm]
\Gamma^{4H}_{ijkl} \!\!\!&= &\!\!\! \sum\limits_{m\le n \le r \le s = 1,2,3}
O_{mi}\, O_{nj}\, O_{rk}\, O_{sl}\ g^{4H}_{mnrs}\,,\qquad
\Gamma^{2H H^+H^-}_{ij} = \sum\limits_{l\le m = 1,2,3}
O_{li}\, O_{mj}\ g^{2H H^+H^-}_{lm}\,,\nonumber
\end{eqnarray}
and
\begin{eqnarray}
  \label{4Hplus}
\Gamma^{4H^+} \!\!\!&=&\!\!\! 
s^4_\beta \lambda_1\: +\: c^4_\beta \lambda_2\:
+\: s^2_\beta c^2_\beta (\lambda_3 + \lambda_4)\: +\:
2s^2_\beta c^2_\beta {\rm Re}\lambda_5\: 
-\: 2s^3_\beta c_\beta {\rm Re}\lambda_6\: -\:
2s_\beta c^3_\beta {\rm Re}\lambda_7\,.\ \qquad 
\end{eqnarray}
The   quantities  $g^{3H}_{ijk}$,   $g^{HH^+H^-}_i$,  $g^{4H}_{ijkl}$,
$g^{2H H^+H^-}_{ij}$ in~(\ref{coupls})  characterize the proper vertex
corrections  to the  corresponding Higgs  self-couplings.   We observe
that, exactly as is the case for $\Gamma^{4H^+}$ in~(\ref{4Hplus}), all
these newly-introduced  couplings depend on  the radiatively-corrected
quartic couplings $\lambda_1$, $\lambda_2$, \dots, $\lambda_7$.  Their
analytic forms are given in Appendix~A.

\section{Constraints on the CPX scenario from LEP searches}
\label{sec3}

The {\sc OPAL} Collaboration~\cite{OPALPN505}  has reported preliminary results
on the search for Higgs bosons in the MSSM with explicit CP violation,
taking   the   parameters   to   be   those   defined   in   the   CPX
scenario~\cite{CEPW},
\begin{eqnarray}
  \label{scenario}
M_{\rm SUSY} \!&=&\! \widetilde{M}_Q\ =\ \widetilde{M}_t\ =\ 
\widetilde{M}_b\ =\ 0.5\ {\rm TeV}\,,\quad
\mu \ =\ 2\ {\rm TeV}\,,\quad 
|A_t| \ =\  |A_b|\ =\ 1\ {\rm TeV}\,, \nonumber\\   
|m_{\tilde{B}}| \!&=&\! |m_{\tilde{W}}|\ =\ 0.2\ {\rm TeV}\,,\qquad
|m_{\tilde{g}}|\ =\ 1\ {\rm TeV}\, ,   
\end{eqnarray}
where $M_{\rm SUSY}$ is the characteristic third generation
squark mass scale, and $m_{\tilde{B}}$, $m_{\tilde{W}}$ are the 
bino and wino masses, respectively.

{\sc OPAL}  presented 
exclusion
regions in  the $M_{H_1}$--$\tan\beta$ plane  for different values of
the CP-violating phases, assuming
$\arg A_{t}=\arg  A_b =
\arg m_{\tilde{g}}=\phi_{\rm  CPX}$, with
$\phi_{\rm CPX}=90^\circ,60^\circ,30^\circ,$  and
$0^\circ$.  Quite generally, there is no reason to expect the equality
of  the phases of the trilinear scalar couplings and the gaugino
masses, nor to assume that $\phi_{\rm CPX}\le 90^\circ$,
and our analysis departs from  this assumption.

To reproduce the results  of the {\sc OPAL} analysis,  we rely on
the combined {\sc ALEPH-DELPHI-L3-OPAL (ADLO)} results for the 
$ZH_{\rm SM}(\to b\bar{b})$
channel ~\cite{LEPHiggs} and  the $hA\to 4b$  channel ~\cite{LHWG0104}.  
Although the experimental $hA$ analysis has been done for
approximately equal values of the masses of the neutral Higgs bosons $h$
and $A$,
there is no major loss in efficiency when the splitting between
these masses becomes larger ~\cite{opal2hdm}, and therefore it is safe to
apply those limits to a more generic set of masses $H_i$, $H_j$ in the
CP-violating scenario.

\begin{figure}[hp]
  \begin{center}
  \resizebox{\textwidth}{!}{
    \includegraphics[bb= 20 20 575 575]{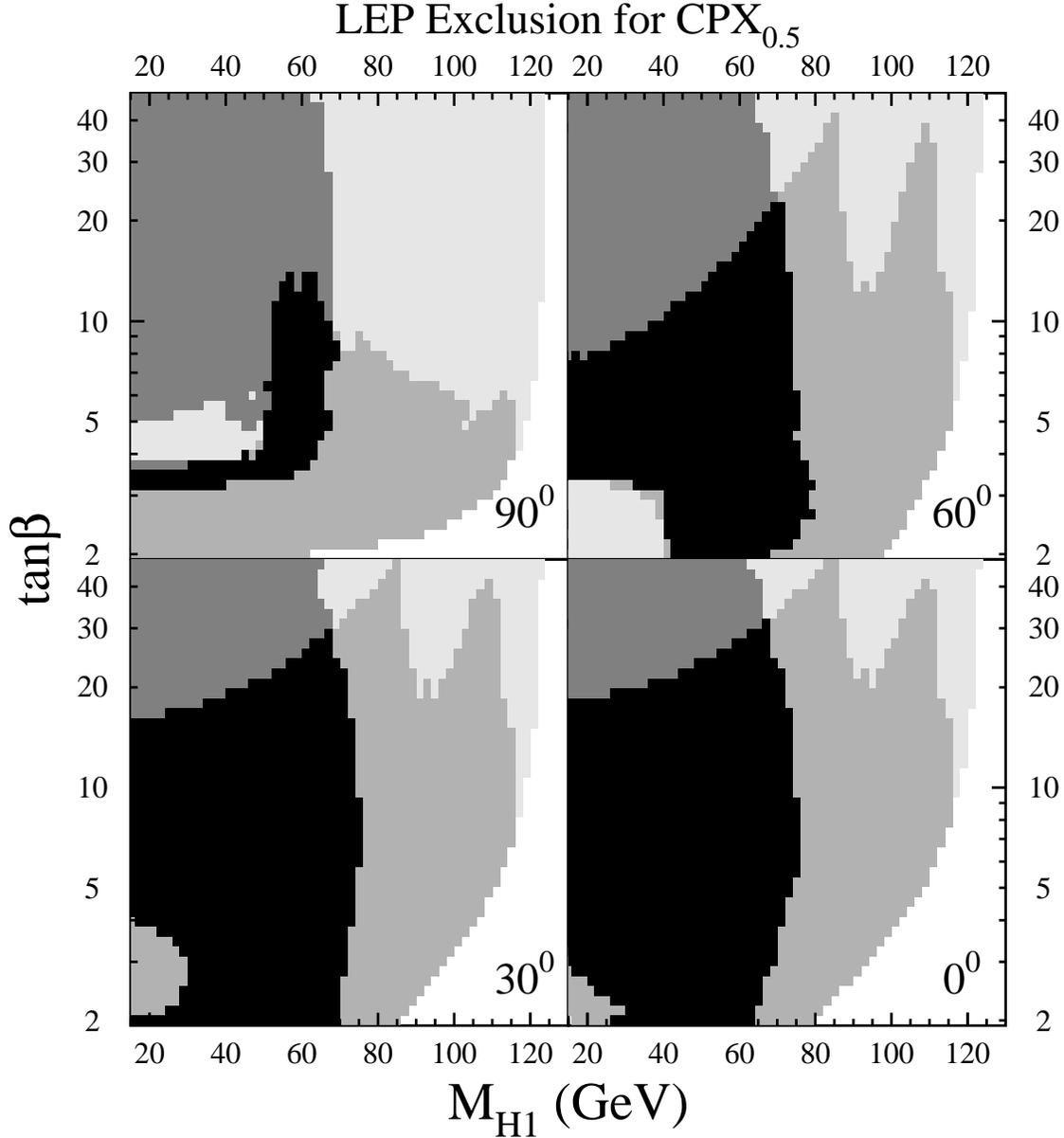}
  }
    \caption{\it Approximate LEP exclusion limits in the 
$M_{H_1}$--$\tan\beta$ 
plane for various CPX scenarios, using combined LEP results. The light
grey covers all the region of parameter space that is consistent with 
electroweak symmetry breaking, the medium grey 
shows the exclusion 
from $e^+ e^- \to Z H_i$, the dark grey shows
the region excluded by $Z^*\to H_i H_j \to 4b$ searches,
and the black region is excluded by both searches.}
\label{fig:opal}
  \end{center}
\end{figure}

Using the  results of these  experimental analyses, we  have generated
Figure~\ref{fig:opal},  based   on  the  CPX   scenario  with  $M_{\rm
SUSY}=0.5$  TeV.   In  this  plot,  the light grey area  covers  the
theoretically  allowed  region  of  parameter space  (consistent  with
electroweak  symmetry  breaking), the  medium grey 
region shows  the
exclusion  from $Z H_i$  final states,  the  dark grey
region is
excluded by the  search for $Z^*\to H_iH_j \to  4b$ final states,
and the black region is excluded by both
searches.

When comparing the results in Fig.~\ref{fig:opal} with the {\sc OPAL} 
results~\cite{OPALPN505}, we
note that  our limits on  $\tan\beta$ are somewhat stronger than
the {\sc OPAL} limits.
This can be attributed  to the fact that
our  analysis is  based  on  combined {\sc ADLO} limits  
obtained  from all  four
experiments~\cite{LHWG0104}.
The results of our analysis show also a somewhat better   
reach than the {\sc OPAL}
one in the $H_i H_j$ production channel. This can 
again be attributed to  our use of the combined results
of all  four  experiments  and the   fact that,  following   the above
discussion,  we have assumed no  deterioration of the $H_i H_j$ signal
for  $M_{H_i}\ne M_{H_j}$.  
We expect that our CPX estimates will be quite similar to the final LEP
combined results.

We see from the cases displayed in Fig.~\ref{fig:opal} that the case of
vanishing phases is most severely constrained by the LEP data.  
Most of the coverage arises from the $ZH_1$ or $H_1H_2$ processes.
The
appearance of two `fingers' in the $ZH_1$ coverage at large $\tan \beta$
for vanishing phases arises from the shape of the LEP exclusion curve,
which is distorted by a marginal excess in the region 
$m_h\sim 80-90$ GeV \cite{LEPHiggs}.  
The case of phase $30^\circ$ is very similar, and the $60^\circ$ case has
analogous features.

For  significant values of the phases,
there are regions  of parameters  in which  the lightest
neutral Higgs boson is very weakly coupled to the $Z$ gauge boson, and
light enough for the heavier neutral Higgs states to decay into a pair
of $H_1$  bosons. In  these regions of  parameters, the  heavier Higgs
states decay into a final state containing four $b$ quarks. Therefore,
the dominant  production and decay modes  contain 6 jets  in the final
state.  The  current  experimental  strategy  is to  employ the
standard 4-$b$ analysis by forcing (with suitable jet definitions) 
the  6-jet
topologies into 4-jet ones.  This leads to a very low efficiency for the
real 6-jet signal, and hence a significant region of CPX parameter
space where light Higgs bosons may exist but are not excluded by the
current analyses.
In our analysis of the MSSM parameter space excluded
by Higgs boson searches at LEP, we have not attempted to simulate the
signatures associated with  the decay of  heavy neutral Higgs
bosons into lighter ones. 
Nonetheless, we are hopeful that a dedicated 6-jet analysis could cover
this region.
Returning to Fig.~\ref{fig:opal}, the   upper    two panels ($90^\circ,
60^\circ$) demonstrate that LEP cannot exclude the presence of a light
Higgs boson at $\tan\beta\sim  3-5, M_{H_1}  \stackrel{<}{{}_\sim} 60$
GeV and $\tan\beta\sim  2-3,  M_{H_1} \stackrel{<}{{}_\sim}  40$  GeV,
respectively, in good  qualitative  agreement with the {\sc OPAL}  results.
Using the {\sc ADLO} data in the $ZH_i(\to b\bar b)$ channel, 
a Higgs boson with the Standard Model coupling can be excluded if $M_{H_i} <
115$ GeV, and even a low-mass Higgs boson can be excluded if 
$|g_{H_iZZ}|\simgt .22$ relative to the Standard Model coupling.

\begin{figure}[!hp]
  \begin{center}
  \includegraphics[width=\textwidth,bb = 20 20 575 575]{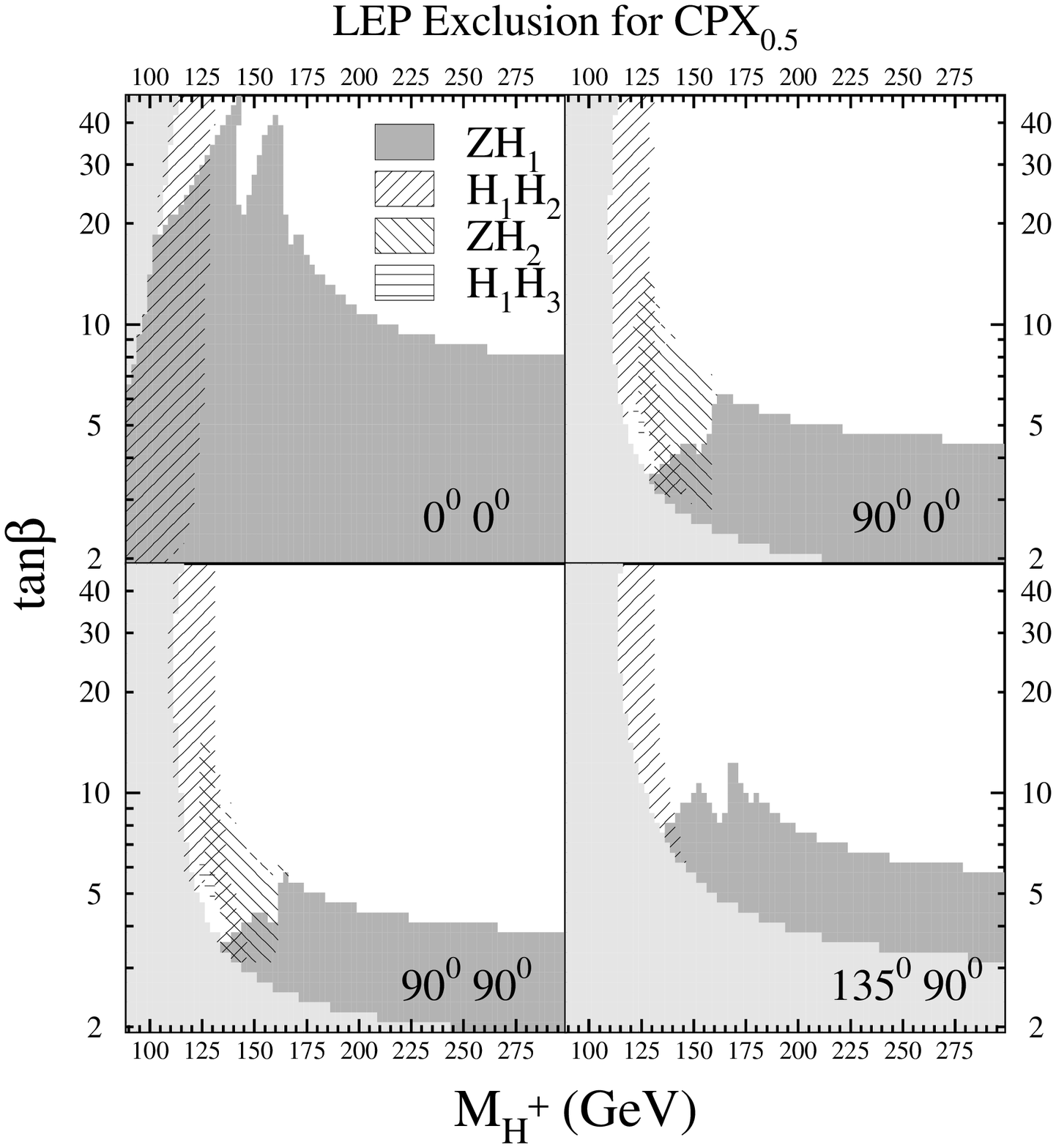}
    \caption{\it LEP exclusion limits in the CPX benchmark scenario
in the $M_{H^+}$--$\tan \beta$ plane for different values of the phases
arg$(A)$ and arg$(m_{\tilde{g}})$ of the trilinear couplings $A_{t,b}$
and the gluino mass parameter, respectively. The four panels show the
results for (0,0); (90,0); (90,90) and (135,90) degrees, respectively. 
The light grey region is disallowed theoretically, the medium grey
region is excluded by the absence of $Z H_1$, $45^\circ$-hatched
region by the absence of $H_1 H_2$, $135^\circ$-hatched region
by the absence of $Z H_2$, and the horizontally-hatched region by the 
absence of $H_1 H_3$.}
    \label{fig:lep1}
  \end{center}
\end{figure}

\begin{figure}[!hp]
  \begin{center}
    \includegraphics[width=\textwidth,bb = 20 20 575 575]{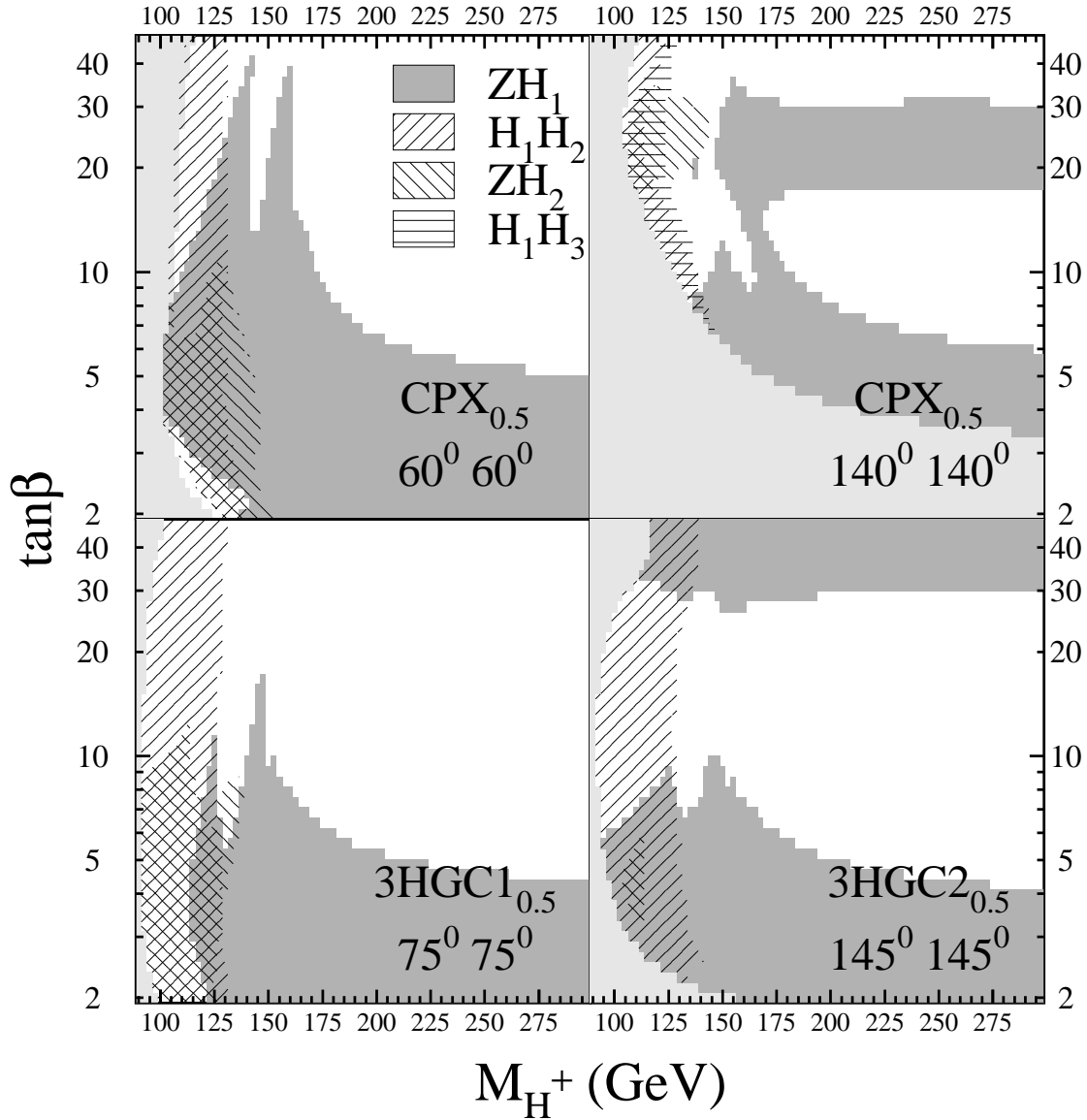}
    \caption{ \it Panels (a) and (b) are similar to
Fig.~\ref{fig:lep1}, but for CP-violating phases (60,60) and (140,140)
degrees. Panels (c) and (d) show two special scenarios 3HGC1 and 3HGC2,
which are particularly challenging for all colliders: 3HGC1: $M_{\rm SUSY}=0.5$
TeV, $\bar A=1.95$, $\bar \mu=2.4$, arg$(A)=\arg(m_{\tilde{g}})=75^\circ$;
3HGC2: $M_{\rm SUSY}=0.5$ TeV, $\bar A=2$, $\bar \mu=2$,
arg$(A)=\arg(m_{\tilde g})=145^\circ$, $M_{\widetilde g}=0.5$ TeV. The 
shadings and 
hatchings have the same significances as in Fig.~\ref{fig:lep1}. }
\label{fig:lep2}
\end{center}
\end{figure}

In Figs.~\ref{fig:lep1} and ~\ref{fig:lep2}, we have presented
a generalization of the above analysis to unequal values
of the CP-violating phases of the trilinear mass parameters
$A_{t,b}$ and the gluino mass. The exclusion plots are presented
in the $M_{H^+}$--$\tan\beta$ plane. For the case of equal
values of the phases, these figures present analogous
information to that presented in Fig.~\ref{fig:opal}.
Note the appearance of fingers for 
arg$(A_{t,b}, m_{\widetilde g})$$=$$(0^\circ,0^\circ)$,
as well as uncovered regions near $\tan\beta\sim 4-5$ and $M_{H^\pm}\sim 
125-140$ GeV for $(90^\circ,90^\circ)$, and
near $\tan\beta\sim 2-3$ and $M_{H^\pm}\sim 105-130$ GeV for
$(60^\circ,60^\circ)$.
Moreover, Figs.~\ref{fig:lep1} and ~\ref{fig:lep2} 
show  distinctively the regions covered by the 
different channels studied at LEP. The light grey regions of 
these figures are excluded theoretically, the medium grey regions 
are excluded by the absence of $Z H_1$, the regions hatched at 
$45^\circ$ by the absence of $H_1 H_2$, the regions hatched at $135^\circ$ 
by the absence of $Z H_2$, and the horizontally-hatched regions by the 
absence of $H_1 H_3$.

Fig.~\ref{fig:lep1} is a study of the interplay of phases
for the trilinear parameters $A_{t,b}$ and the gluino mass
parameter.  The appearance
of substantial phases to the  parameters can significantly modify
the collider phenomenology (the phase of $A_t$ providing the dominant
effect). The addition of a phase to the gluino mass parameter,
however contributes to CP-violating effects only at the two-loop level
and its effect is comparatively smaller, as can be seen
in Fig.~\ref{fig:lep1}. For the case of arg$(A_{t,b}) = 90^\circ$,
it is clear from Fig.~\ref{fig:lep1} that, for $\tan\beta \simeq 4$--5,
as the Higgs $H_1$ becomes lighter, it starts decoupling from the 
$Z$. There is a significant region of parameter space where, although
heavier, $H_2$ is in the kinematic region accessible to LEP and couples
relevantly to the $Z$ gauge boson. For $M_{H^+} \simlt 130$ GeV,
$H_1$ becomes light enough that the decay $H_2 \to H_1H_1$
dominates over $H_2 \to b\bar{b}$, and therefore $H_2$ detection becomes
difficult. Finally, for arg$(A_{t,b}) = 135^\circ$ and 
arg$(m_{\tilde{g}}) = 90^\circ$, and for moderate or large values
of the charged Higgs mass, all neutral Higgs bosons rapidly become
 sufficiently heavy to be out of the reach of LEP for moderate
and large values of $\tan\beta$.

In Fig.~\ref{fig:lep2} we display results for the CPX scenario,
for equal values of the $A_{t,b}$ and gluino mass phases equal
to $60^\circ$ and $140^\circ$. The region of parameters left uncovered
by LEP for low values of $\tan\beta$ and of the charged Higgs 
boson mass, for phases equal to $60^\circ$,
is seen more clearly than in Fig.~\ref{fig:opal}. It
is also apparent that, as happens for a CP-violating phase
equal to $90^\circ$, close to the region left uncovered by LEP, there
is a large region of parameters covered by the $ZH_2$ channel.

\begin{figure}
\begin{center}
\includegraphics[scale=0.5,bb= 0 0 567 425]{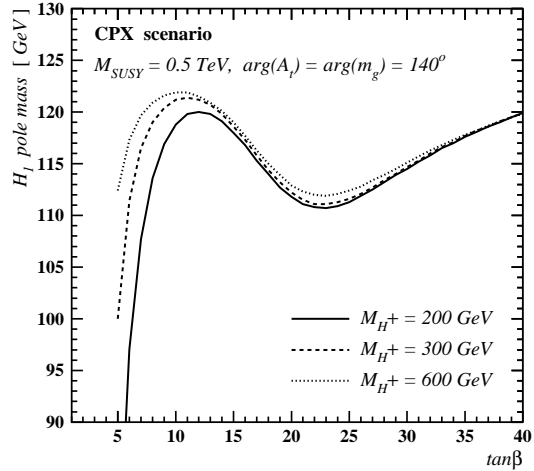}
\caption[]{\it The non-monotonic behavior of $M_{H_1}$, the mass of the 
lightest neutral Higgs boson, in the CPX scenario with arg($A_t$) = 
arg($m_{\widetilde g}$) = 140$^\circ$ for different values of $M_{H^+}$. 
The 
dip for $15 < \tan \beta < 30$ is due to the behavior of the 
radiative corrections from the sbottom sector, which are screened at 
large $\tan \beta$, as discussed in the text.} 
\label{fig:updownup}
\end{center}
\end{figure}

For phases equal to $140^\circ$, we find a peculiar behavior of the
covered region of parameters for large values of the charged Higgs mass,
which can be traced to the behavior of the $H_1$ mass shown in
Fig.~\ref{fig:updownup}. For small and moderate values of $\tan\beta$, it
has the standard behavior of the MSSM without explicit CP violation, 
increasing with $\tan\beta$. For $\tan\beta $
above about 10, however, it starts decreasing due to the effect of
radiative corrections from the sbottom sector, which involve the bottom
Yukawa coupling. The bottom quark Yukawa corrections are
screened for sufficiently large values of
$\tan\beta$, causing the mass of the $H_1$ to increase again and
eventually become larger than the LEP kinematic limit.

This screening phenomena occurs because the sbottom--induced radiative
corrections are complex.
At large values of $\tan\beta$,
the bottom-quark 
Yukawa coupling $h_b$ is approximately related to the bottom mass and to
the Yukawa corrections $\Delta h_b$ by
\begin{equation}
h_b = \frac{m_b}{v} \; \frac{\tan\beta}{1 + (\Delta h_b/h_b) \tan\beta}
\end{equation}
The negative corrections to the Higgs mass squared coming from sbottom
loops are  proportional to the fourth  power of the  modulus of $h_b$.
For equal values of the phases of the gluino and of the trilinear mass
parameter $A_t$, i.e.\ ${\rm arg}  (A_t) = {\rm arg} (m_{\tilde{g}}) =
\phi_{\rm CPX}$, the modulus of $h_b$ is given by
\begin{equation}
|h_b| = \frac{m_b}{v} \; \frac{\tan\beta}
{\sqrt{ 1 + 
2 |\Delta h_b/h_b| \cos(\phi_{\rm CPX}) \tan\beta 
+ \left(|\Delta h_b/h_b| \tan\beta \right)^2}}
\end{equation}
For  $\cos(\phi_{\rm  CPX})  \geq   0$,  the  above  expression  is  a
monotonically  increasing function  of $\tan\beta$,  meaning  that the
negative corrections induced by sbottom loops become larger for larger
values of $\tan\beta$.

On the other hand, for  negative values of $\cos(\phi_{\rm CPX})$, the
expression for $|h_b|$ has a maximum at a value of
\begin{equation}
\tan\beta^{\rm max} = -\frac{1}{|\Delta h_b/h_b| \cos(\phi_{\rm CPX})}
\end{equation}
meaning that the negative effects on the Higgs mass induced by sbottom
loops are  most pronounced at intermediate values  of $\tan\beta$, and
are  reduced again  at large  $\tan \beta$.   For the  particular case
under  consideration,  we  have  $|\Delta h_b|/h_b  \simeq  1/20$  and
$\phi_{\rm  CPX}  =  140^\circ$,  with  $\cos(\phi_{\rm  CPX})  \simeq
1/\sqrt{2}$, while  $|h_b| \stackrel{<}{{}_\sim} 1$  stays well within
the  perturbative range.  Thus, lower  values  of the  Higgs mass  are
obtained for $\tan\beta  \simeq 25$ than for either  smaller or larger
values   of   $\tan\beta$.   This   feature   is   seen   clearly   in
Fig.~\ref{fig:updownup}, and explains the  stronger coverage of LEP at
this  intermediate  region  of   $\tan\beta$, as seen  in  panel  (b)  of
Fig.~\ref{fig:lep2}.

Apart from the above property, the only salient feature is that in the
region between the one covered by LEP in the $ZH_1$ channel and that
covered by LEP in the $H_1H_2$ channel, $H_1$ becomes light but
couples very weakly to the $Z$ boson. 

The last two scenarios considered in Fig.~\ref{fig:lep2} are
interesting since, for moderate values of the charged Higgs mass of
about 150 GeV, the three neutral Higgs bosons become quite similar in
mass and share approximately equal couplings to the weak gauge
bosons.  We call these tri-Higgs-gauge coupling (3HGC)
scenarios, which are denoted in short as 3HGC1 and 3HGC2. As we 
discuss below, the searches at hadron colliders become particularly
difficult in 3HGC1 and 3HGC2.  The excluded region for moderate and
large values of the charged Higgs-boson mass and large values of $\tan\beta$
in the scenario 3HGC2 is related to the Higgs-sbottom coupling effect
discussed above.

To summarize the results from LEP searches, the phenomenology is
mainly sensitive to the values of $g_{H_iZZ}$ and the kinematic limit
on the Higgs boson masses.  The exceptions are the cases when the
decay $H_2\to H_1 H_1$ becomes relevant at small $\tan\beta$ and when
radiative corrections extend the LEP coverage for certain choices of
phases $\phi_{\rm CPX}$ at large $\tan\beta$.
A Higgs boson with the Standard Model coupling can be excluded if $M_{H_i} <
115$ GeV, and even a low-mass Higgs boson can be excluded if 
$|g_{H_iZZ}| \simgt .22$ relative to the Standard Model coupling.
Note that the application of the $ZH_i$ and $H_i H_j$
results to the CPX scenarios treats each potential signature
independently.  Any possible complications from one potential signal
effecting the background estimate for another (and {\it vice versa})
are ignored.  We are not aware of any experimental analysis of such a
situation.

\section{Higgs Boson Searches at Hadron Colliders}
\label{sec4}

Experiments at the Tevatron and LHC will probe the MSSM Higgs sector
beyond the kinematic and dynamical reach of LEP.
At these hadron colliders, we have considered
several channels for Higgs boson $H_i$ searches, where $H_i$ stands for
$H_{1,2,3}$:
\begin{description}
\item{(a)} $t\bar t H_i (\to b\bar b)$ at the LHC
\item{(b)} $W/ZH_i (\to b\bar b)$ at the Tevatron
\item{(c)} $WW\to H_i (\to \tau^+\tau^-)$ at the LHC
\item{(d)} $gg\to H_i \to \gamma\gamma$ at the LHC.
\end{description}
In the
absence of reliable experimental simulations, as in the case of LEP,
we have not modified our analysis to account for the decay $H_j\to
H_iH_i\to 4b$ $(j>i)$, should it become relevant.
The search channels $(a)-(d)$ are considered to be the most promising for 
observing a Standard Model Higgs boson, and have been the most thoroughly
analyzed and simulated.  
Since the MSSM Higgs sector has a decoupling limit, where there is
a light Higgs boson with properties almost indistinguishable from
a SM Higgs boson, it is appropriate to apply these SM analyses to
the MSSM case.
However, one would have to be careful in 
interpreting a signal in some of these channels.  For example, 
signature $(d)$ may arise when $H_i$ is replaced by a pseudo-Goldstone
boson, as in Technicolor models.  Signature $(a)$ may arise, for example,
when $H_i$ is replaced by a top-pion, as in a Topcolor model.  
Only observation of channels
$(b)$ and $(c)$ would clearly indicate that the scalar $H_i$ is
associated with EWSB.  Signature $(a)$ may indicate that the scalar 
responsible for EWSB also generates fermion masses.  Nonetheless, if
signature $(a)$ or $(d)$ were observed with the rate expected in a
weakly--coupled theory, it could be considered strong evidence for
a fundamental Higgs scalar or scalars.
In the present work, we do not analyze those signatures associated
with non--SM--like Higgs bosons, such as charged Higgs or pseudoscalar
Higgs bosons, since their interpretation would be ambiguous.

\begin{figure}[!hp]
  \begin{center}
  \resizebox{\textwidth}{!}{
    \includegraphics[bb= 20 20 575 575]{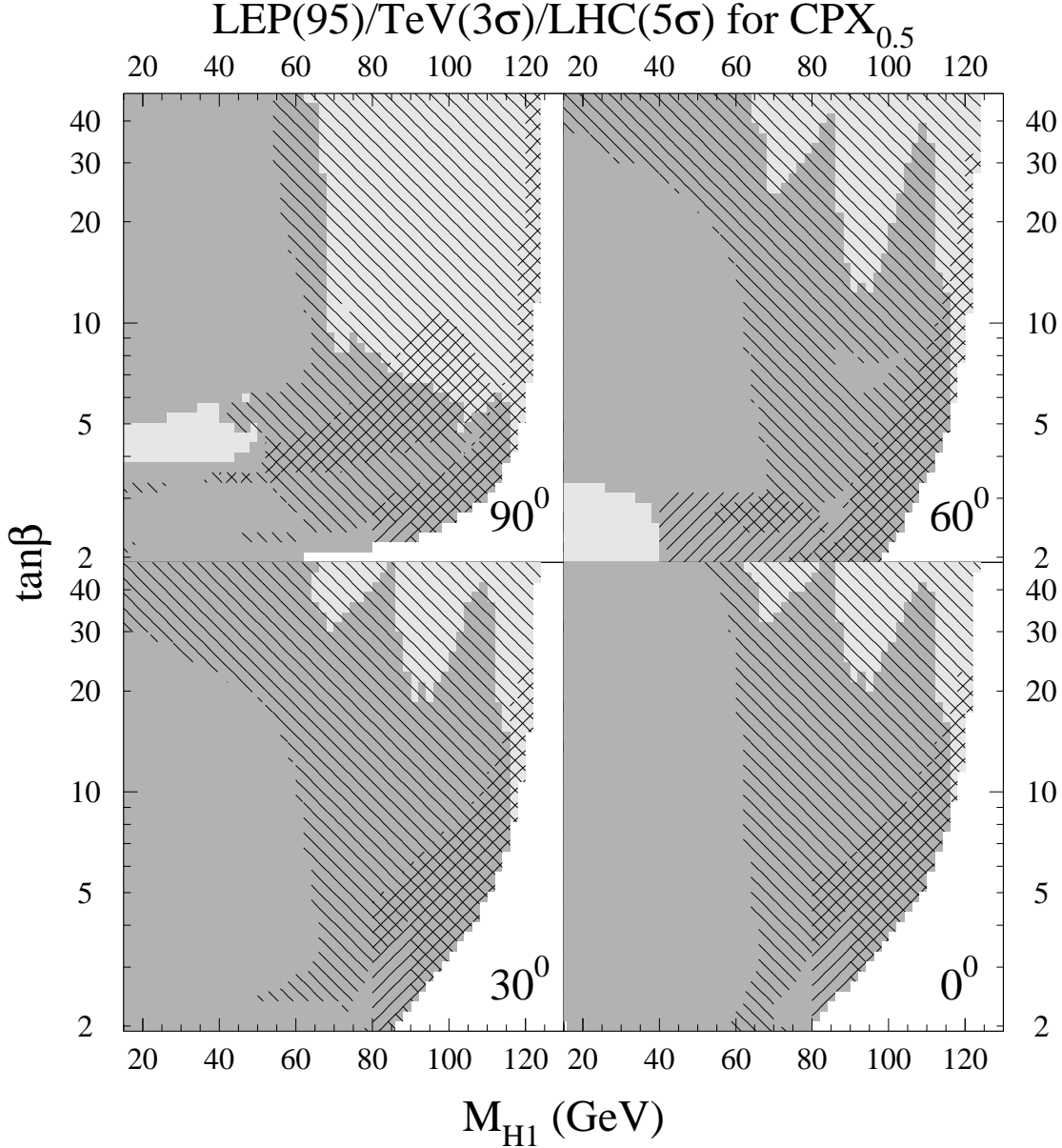}
  }
    \caption{\it Approximate Tevatron/LHC discovery and 
LEP exclusion limits in the $M_{H_1}$--$\tan\beta$ plane for the CPX 
scenario with both phases set to: (a) $90^\circ$, (b) $60^\circ$,
(c) $30^\circ$, and (d) $0^\circ$.
The reach of the Tevatron $W/ZH_i(\to 
b\bar b)$ search is shown as $45^\circ$ lines and that of the combined
LHC search channels as $135^\circ$ lines.The combined LEP 
exclusion is shown in medium gray, superimposed on the
theoretically allowed region in light grey.}
    \label{fig:lhc_opal}
  \end{center}
\end{figure}

Figure~\ref{fig:lhc_opal} shows the coverage of the 
$M_{H_1}$--$\tan\beta$
plane by the Tevatron $W/ZH_i(\to b\bar b)$ search ($45^\circ$ lines) and 
the combined LHC coverage for the $t\bar tH_i(\to b\bar b)$, $WW\to
H_i(\to\tau^+\tau^-)$ and
$gg\to H_i(\to\gamma\gamma)$ searches 
($135^\circ$ lines) for the same values of the 
CP-violating
phases as chosen in Fig.~\ref{fig:opal}.  
For the Tevatron, we show the $3\sigma$ evidence coverage with 5 fb$^{-1}$,
while, for the LHC, we show the $5\sigma$ discovery coverage for the 
$\gamma\gamma$ and $b\bar b$ channels
with 100 fb$^{-1}$ and the $\tau^+\tau^-$ channel with 30 fb$^{-1}$.
 The previous LEP 95\% C.L. exclusion is also
included (medium grey) superimposed on the
theoretically allowed region (light grey).  
While the Tevatron and LHC searches 
are adequate for
extending the coverage into the large-$\tan\beta$ region, the region of
small $M_{H_1}$ and low $\tan\beta$ for $\phi_{\rm CPX}=90^\circ,60^\circ$
remains uncovered. Although the persistence of these regions can be
clearly traced to the decay of the heavier Higgs bosons into
lighter ones, one could inquire why the LHC cannot see the {\em lighter}
Higgs boson in the $t\bar t H_1$ and/or the gluon fusion channels. The
reason is that, in the same region of parameters, $H_1$ couples
weakly to the top quark, as well as negligibly to $W$ and $Z$ bosons.
Therefore, not only is the $t\bar{t}H_1$ production channel suppressed
but the loop-induced gluon fusion production and decay into photons
is suppressed as well.  Furthermore, since $\tan\beta$ is small, $H_1$
would likely not be observed in the pseudoscalar Higgs channel
$b\bar b H_1(\to b\bar b)$.

If, contrary to our expectations, a dedicated LEP analysis cannot
cover this problematic region, there still remains the possibility
of observing $H_2\to H_1 H_1$ decays (when $H_2$ is otherwise
SM--like) at hadron colliders.  Since $H_2$ is SM--like, there still
is a substantial production cross section.  Dedicated searches 
{\em should} be able to identify $4b$ final states with high
efficiency and substantially less
background than for the $2b$ case.  
Various cases of Higgs pair production, such
as $gg\to H\to hh/AA$, have been considered in previous
studies for the Tevatron and LHC, but mostly at the parton
level \cite{Dai:1995cb,Atlas}.  
The studies would
have to be updated for the case of explicit CP violation.
Searches for $W/ZH_2(\to 4b)$ or $t\bar t H_2(\to 4b)$ may be 
more promising, if only because of the reduced backgrounds
from requiring the leptonic decay of a $W$ boson.

Outside these small regions of parameters, the search for neutral
Higgs bosons at the Tevatron and LHC colliders are complementary and
sufficient to cover the full parameter space in one or several
channels.  The Tevatron does not provide any additional coverage alone
over all three LHC search channels.   However, if we just focus on the
$W/ZH_i$ and $WW\to H_i$ channels  -- the only two which are dependent
on  the coupling  $g_{H_iWW}$ --  then both  are nearly  sufficient to
cover the entire parameter space: for $90^\circ$, $M_{H_1}\sim 80$ GeV
and  $\tan\beta\sim 10$,  a  small region  is  left uncovered  without
$t\bar t H_i$.  This complementarity reflects the different production
and decay channels that are being used in the search for neutral Higgs
bosons,   and    is   similar    to   that   in    the   CP-conserving
scenario~\cite{CMW}.  For  instance, there  are  regions of  parameter
space  where  either  the   $b\bar{b}$  or  the  $\tau^+\tau^-$  decay
branching ratio of the lightest neutral Higgs boson is suppressed with
respect to the Standard Model ones, while the other neutral Higgs
bosons are heavy and weakly coupled to the gauge bosons. These
suppressions tend to occur for large values of $\tan\beta$. 
The large values of $\mu$ chosen in our examples maximize the
CP-violating effect, but also determine that
for large values of $\tan\beta$ the quantum corrections to the
bottom Yukawa coupling are large and sizeable, causing a visible
displacement of the regions of suppressions of each coupling.
Since, for a light enough Higgs, these two branching
ratios tend to be very important, a suppression of one of them tends
to enhance the other. We discuss specific examples below.
For smaller values of $\mu$, both the $b\bar b$ and
$\tau^+\tau^-$ couplings can be suppressed 
simultaneously and then the $\gamma
\gamma$ decay branching ratio of the lightest Higgs boson tends to be
enhanced.  An analysis
of these cases can be found in Ref.~\cite{CMW}.

\subsection{Detailed Analysis of the different scenarios}

\begin{figure}[!hp]
  \begin{center}
   \includegraphics[width=\textwidth,bb= 20 20 575 575]{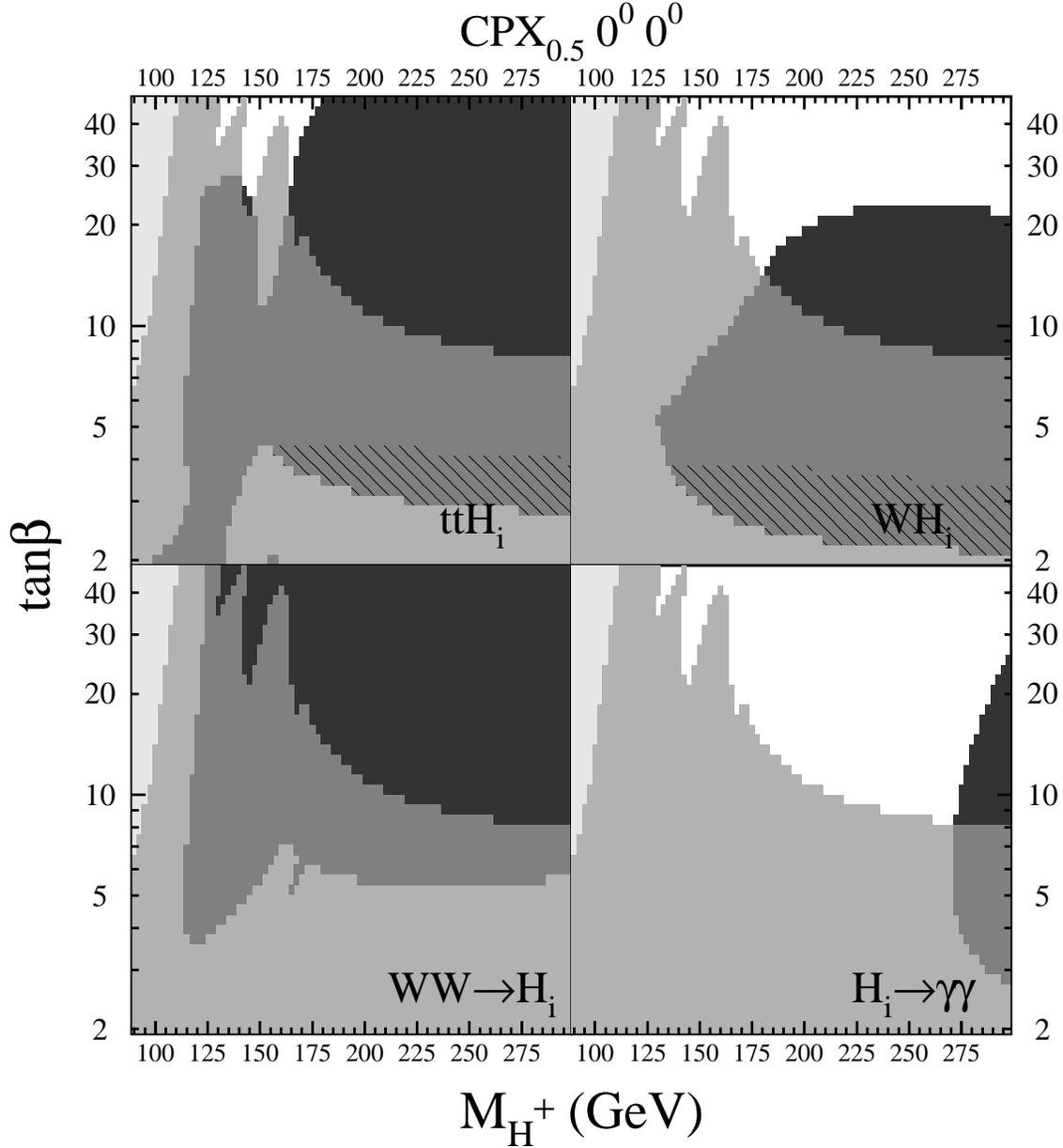}
\caption{\it Coverage of the $M_{H^\pm}$--$\tan\beta$ plane in the CPX
scenario with $M_{\rm SUSY}=0.5$ TeV and for the phases shown at top.
The light grey region is
theoretically excluded, and the two medium grey regions are excluded by
LEP~.  The two darkest shades of grey 
are (a) the 5-$\sigma$
discovery region at the LHC using $t\bar t H_i\to b\bar b$ (100
fb$^{-1}$); (b) the 3-$\sigma$ evidence region at the Tevatron using
$W/Z H_i\to b\bar b$ (5 fb$^{-1}$); (c) the 5-$\sigma$ discovery
region at the LHC using $WW\to H_i \to\tau^+\tau^-$ (30 fb$^{-1}$);
and (d) the 5-$\sigma$ discovery region at the LHC using $H_i \to
\gamma\gamma$ with 100 fb$^{-1}$ of luminosity.  The $135^\circ$
lines in panels (a), (b),  and (d) show the coverage for
30, 2, and 30 fb$^{-1}$ of data, respectively.  This baseline model
shows the behavior when no CP violation is present, i.e., the phases
of $A_t=A_b$ and $m_{\widetilde g}$ are both zero.  }
\label{com00}
  \end{center}
\end{figure}

\begin{figure}[!hp]
  \begin{center}
\includegraphics[width=\textwidth,bb = 20 20 575 575]{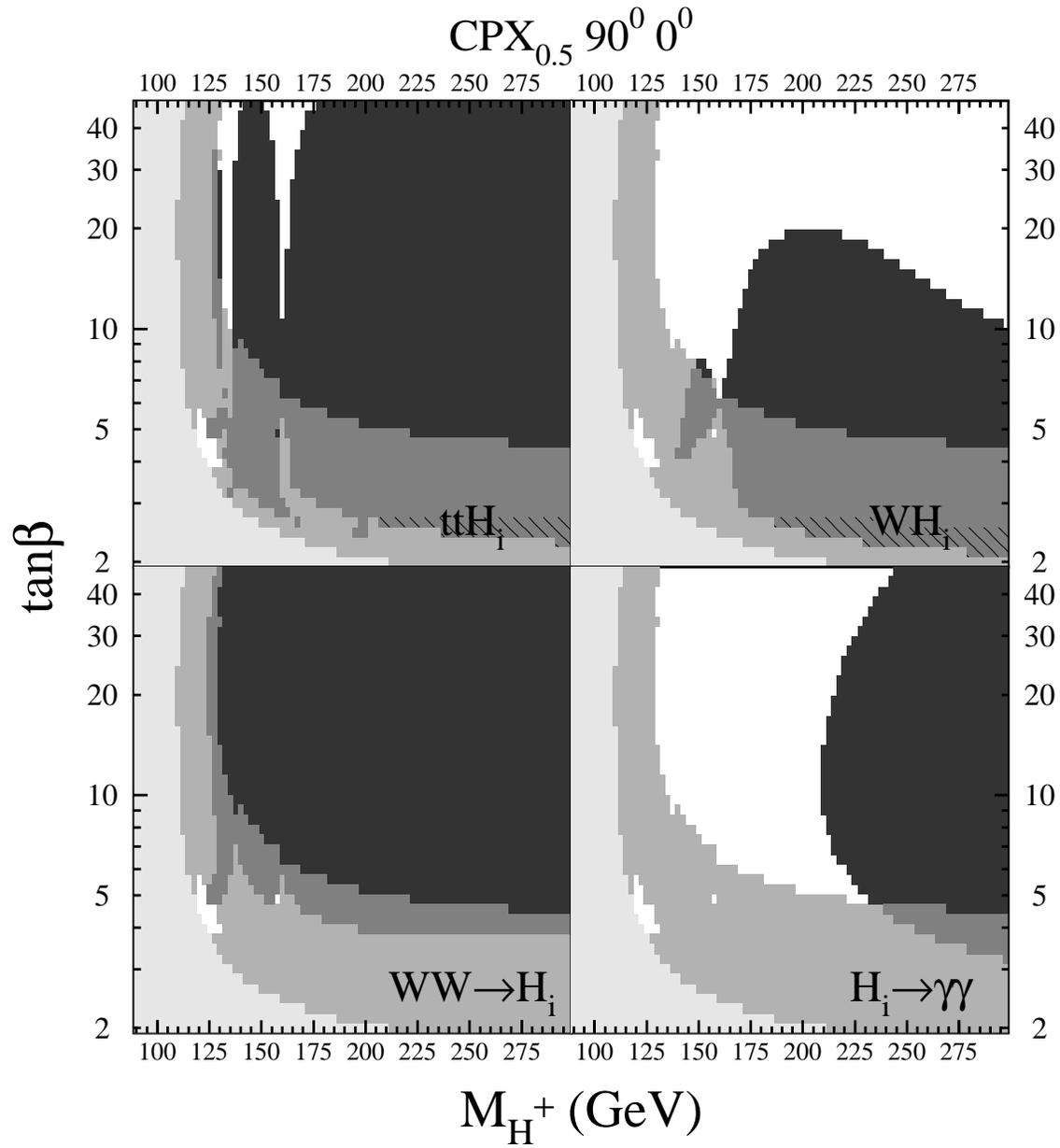}
\caption{\it Same as Fig.~\ref{com00}, but for CP-violating
phases (arg($A_{t,b}$), arg($m_{\tilde{g}}$)) = ($90^\circ,0^\circ$).}
\label{com900}
  \end{center}
\end{figure}

\begin{figure}[!hp]
  \begin{center}
\includegraphics[width=\textwidth,bb= 20 20 575 575]{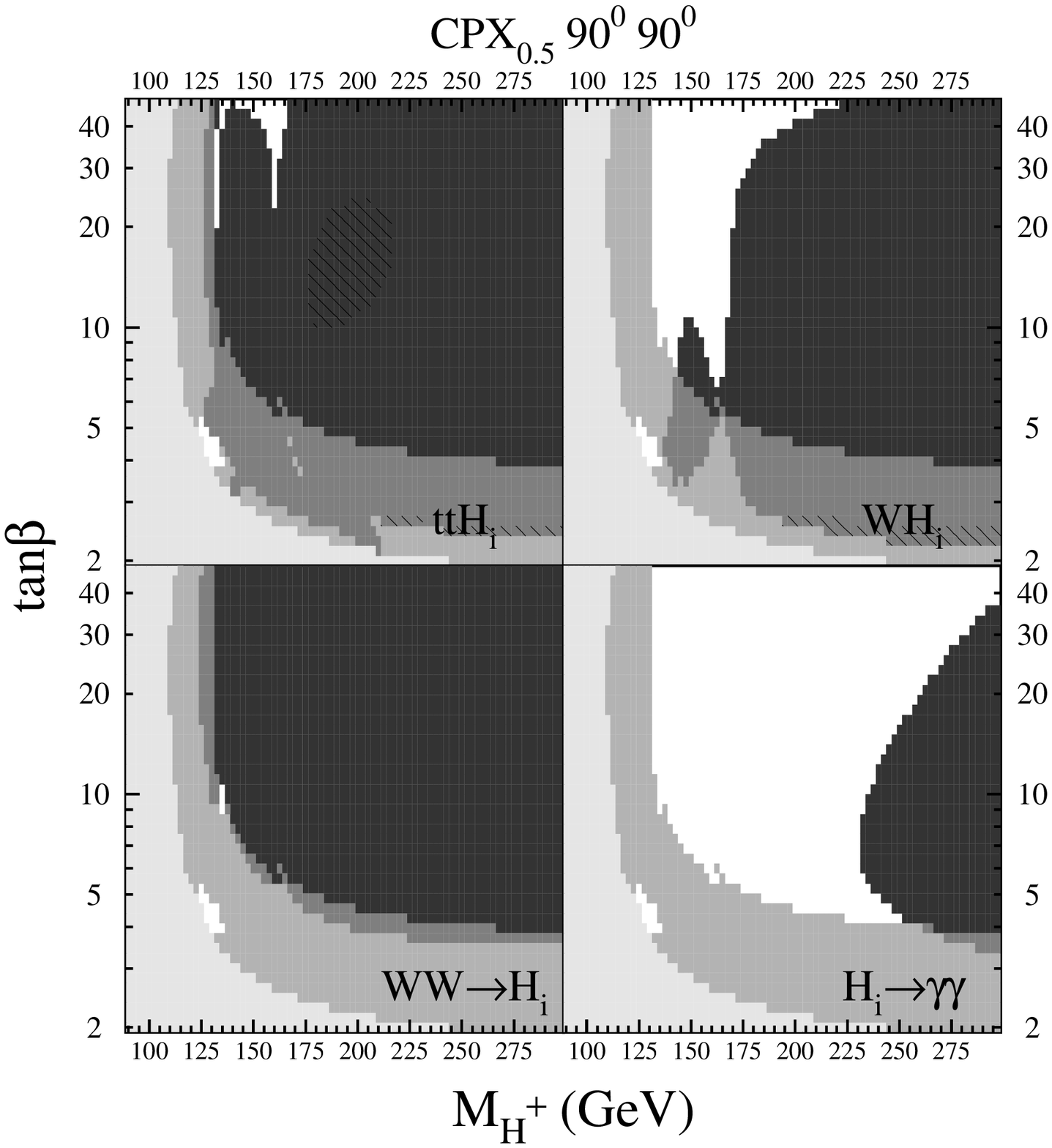}
\caption{\it Same as Fig.~\ref{com00}, but for CP-violating
phases (arg($A_{t,b}$), arg($m_{\tilde{g}}$)) = ($90^\circ,90^\circ$).}
\label{com9090}
  \end{center}
\end{figure}

\begin{figure}[!hp]
  \begin{center}
\includegraphics[width=\textwidth,bb = 20 20 575 575]{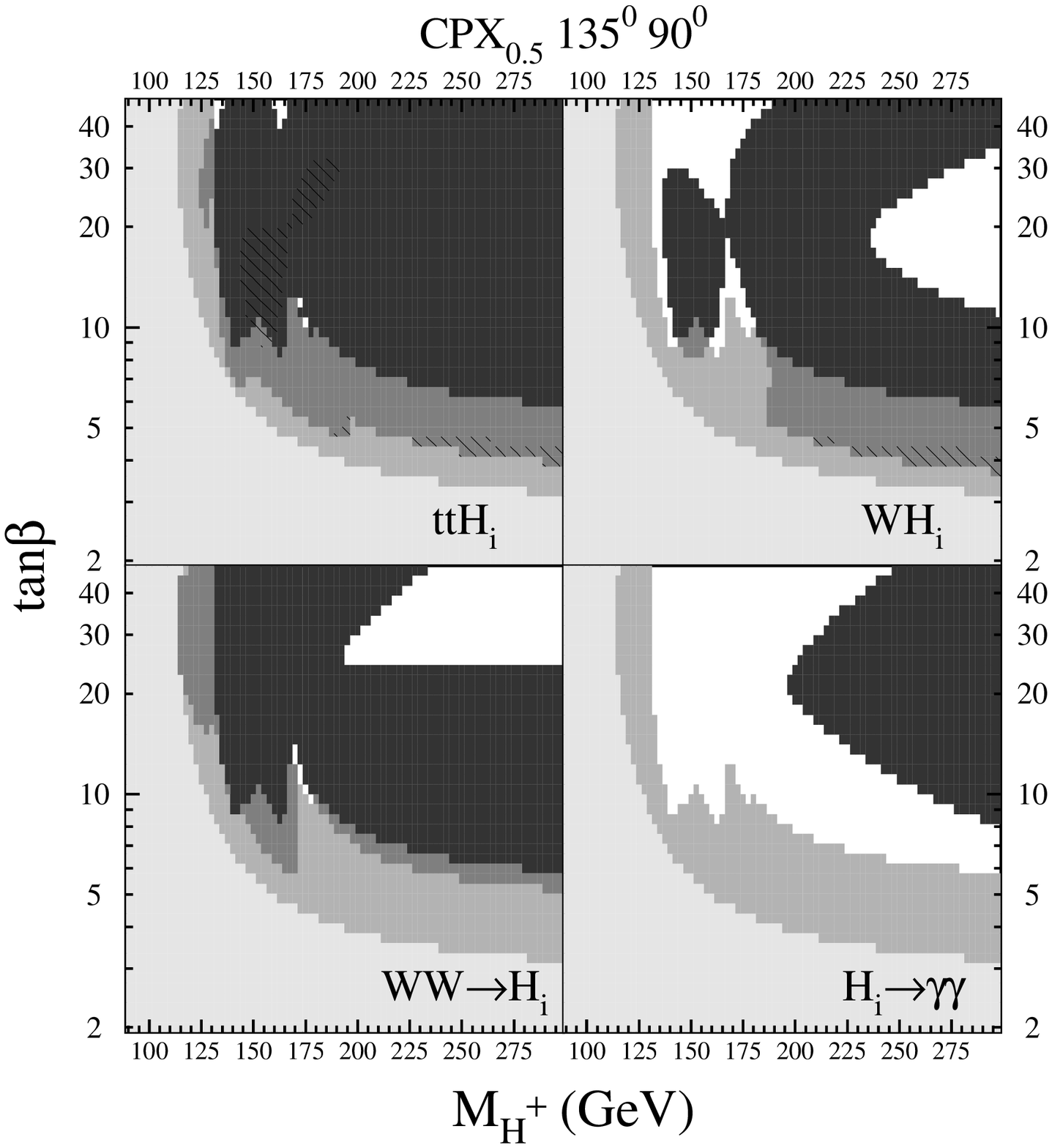}
\caption{\it Same as Fig.~\ref{com00}, but for CP-violating
phases (arg($A_{t,b}$), arg($m_{\tilde{g}}$)) = ($135^\circ,90^\circ$).}
\label{com13590}
  \end{center}
\end{figure}

\begin{figure}[!hp]
  \begin{center}
\includegraphics[width=\textwidth,bb= 20 20 575 575]{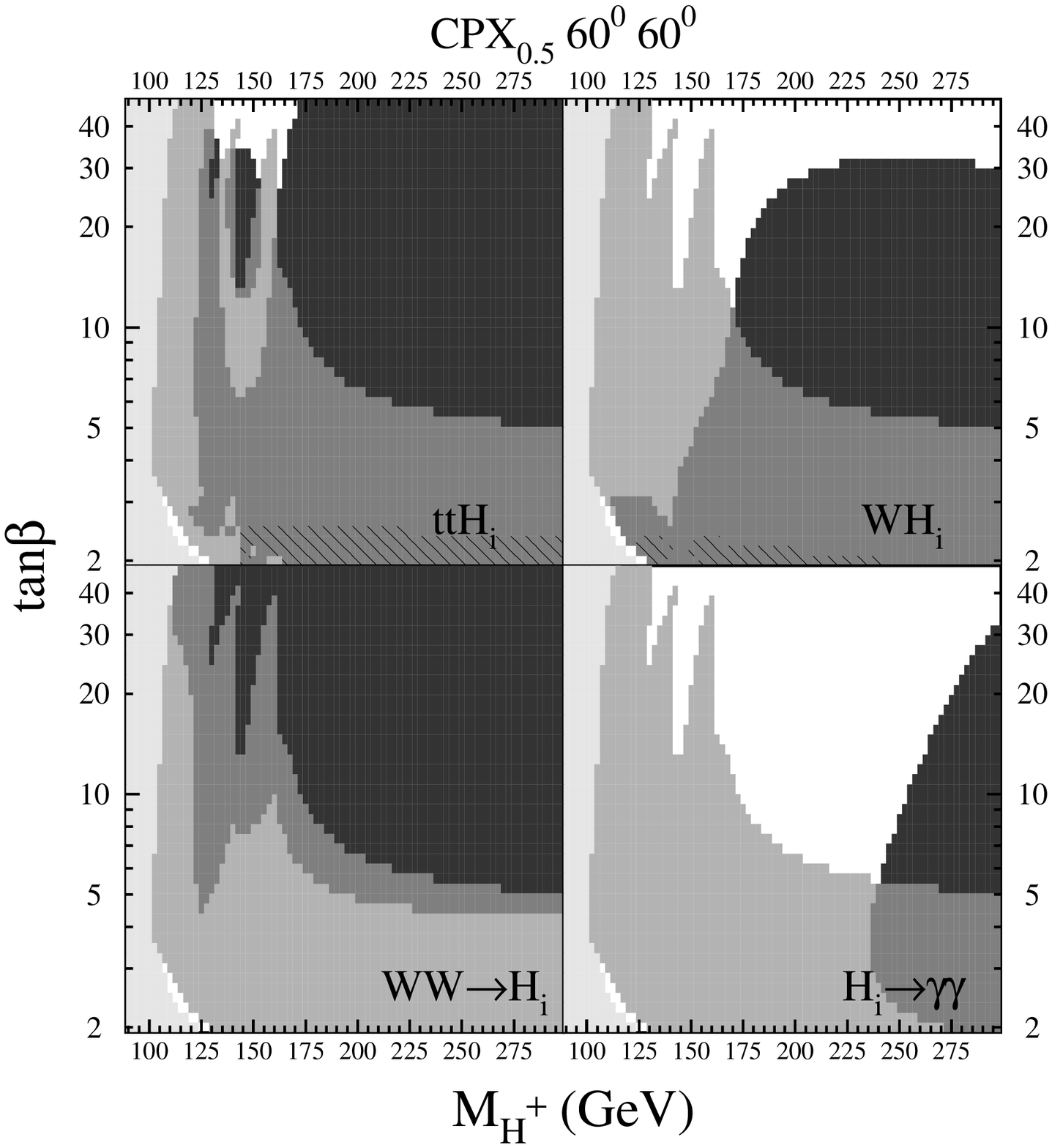}
\caption{\it Same as Fig.~\ref{com00}, but for CP-violating
phases (arg($A_{t,b}$), arg($m_{\tilde{g}}$)) = ($60^\circ,60^\circ$).}
\label{com6060}
  \end{center}
\end{figure}

\begin{figure}[!hp]
  \begin{center}
\includegraphics[width=\textwidth,bb= 20 20 575 575]{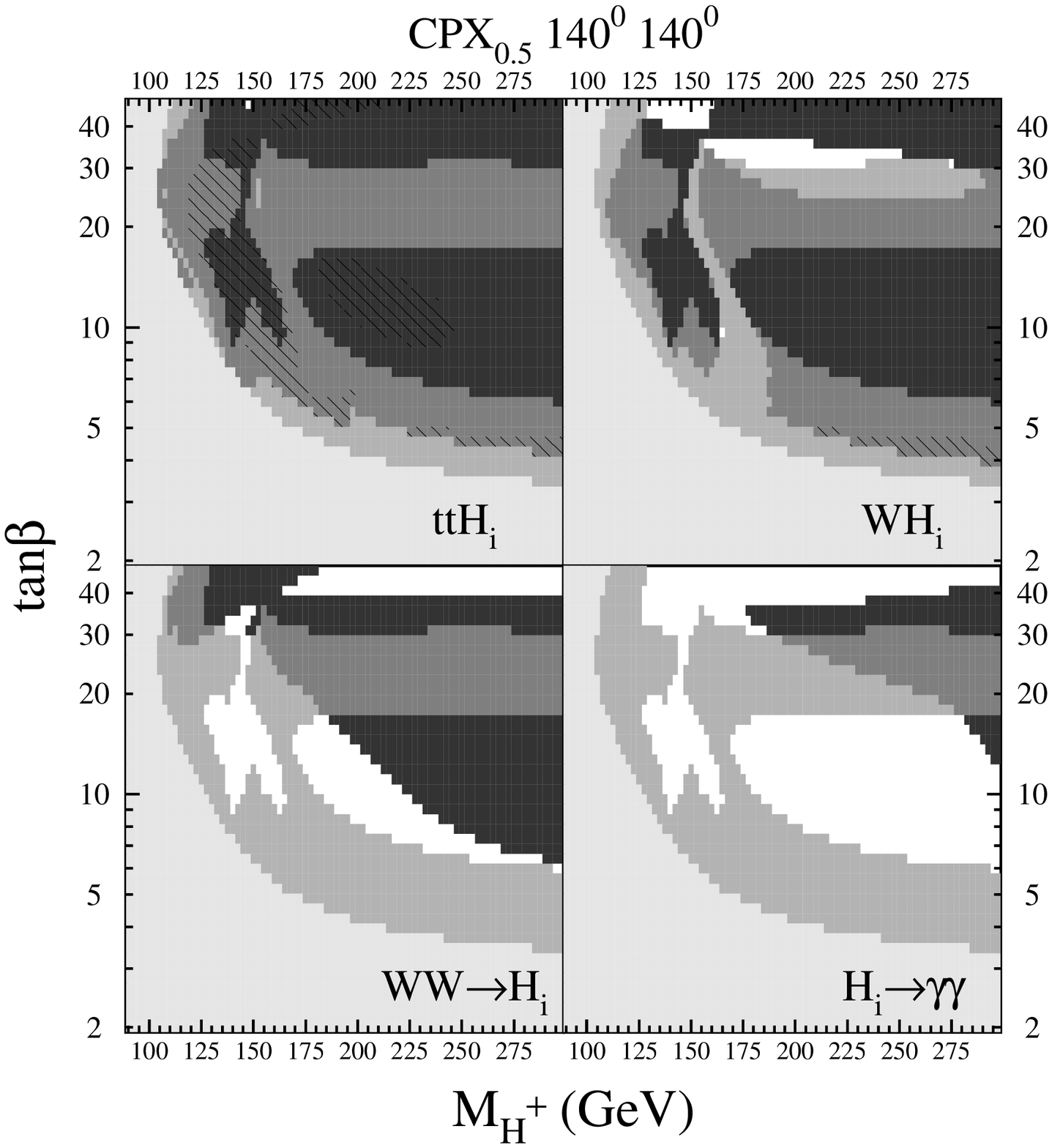}
\caption{\it Same as Fig.~\ref{com00}, but for CP-violating
phases (arg($A_{t,b}$), arg($m_{\tilde{g}}$)) = ($140^\circ,140^\circ$).}
\label{com140140}
  \end{center}
\end{figure}

\begin{figure}[!hp]
  \begin{center}
\includegraphics[width=\textwidth,bb= 20 20 575 575]{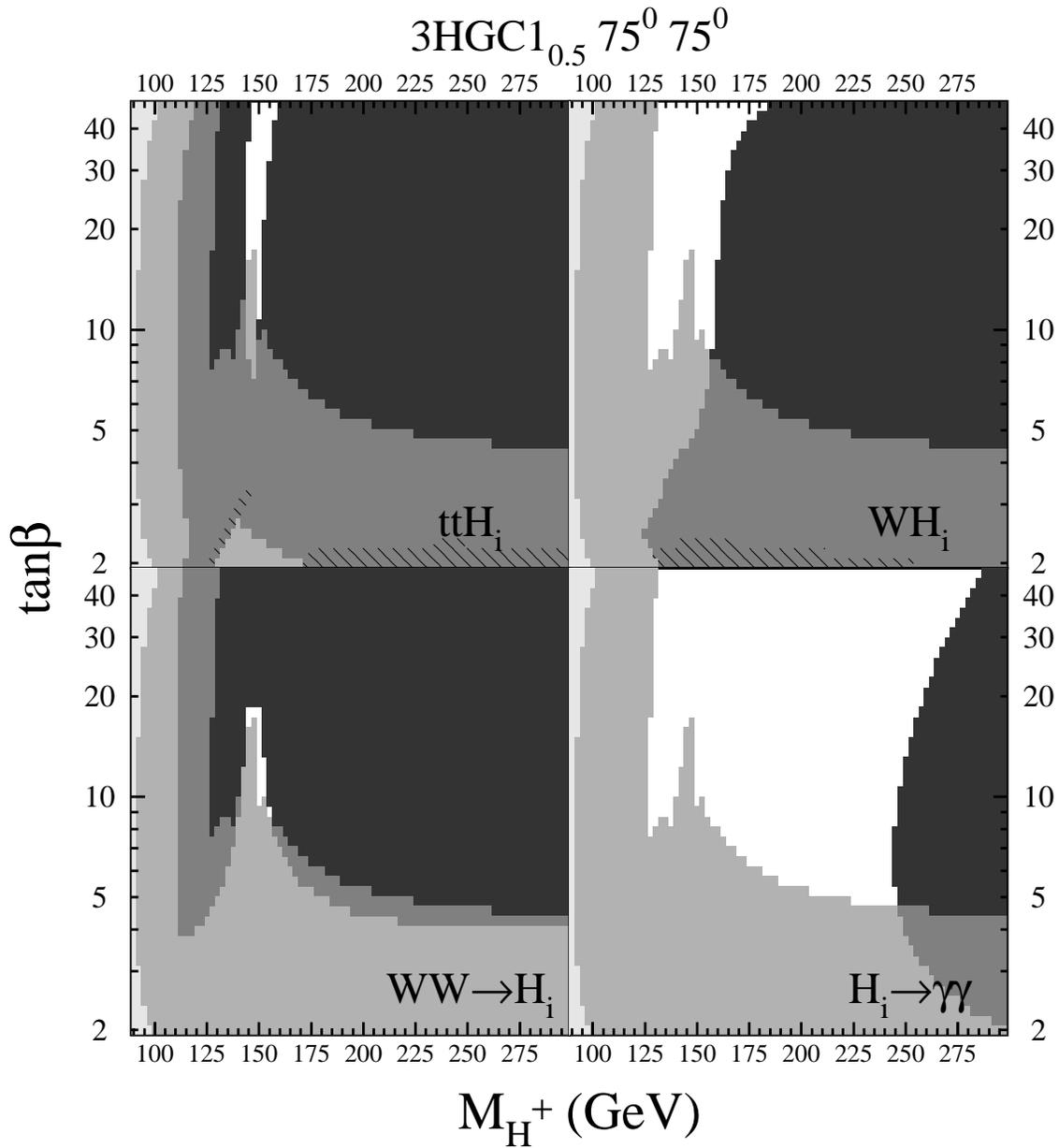}
\caption{\it Same as Fig.~\ref{com00}, but for the 3HGC1 scenario
defined in Fig.~\ref{fig:lep2}.  Here, a region remains uncovered due
to dilution of the $g_{H_iVV}$ ($V=W,Z$) coupling among three Higgs bosons.}
\label{comtry}
  \end{center}
\end{figure}

\begin{figure}[!hp]
  \begin{center}
\includegraphics[width=\textwidth,bb= 20 20 575 575]{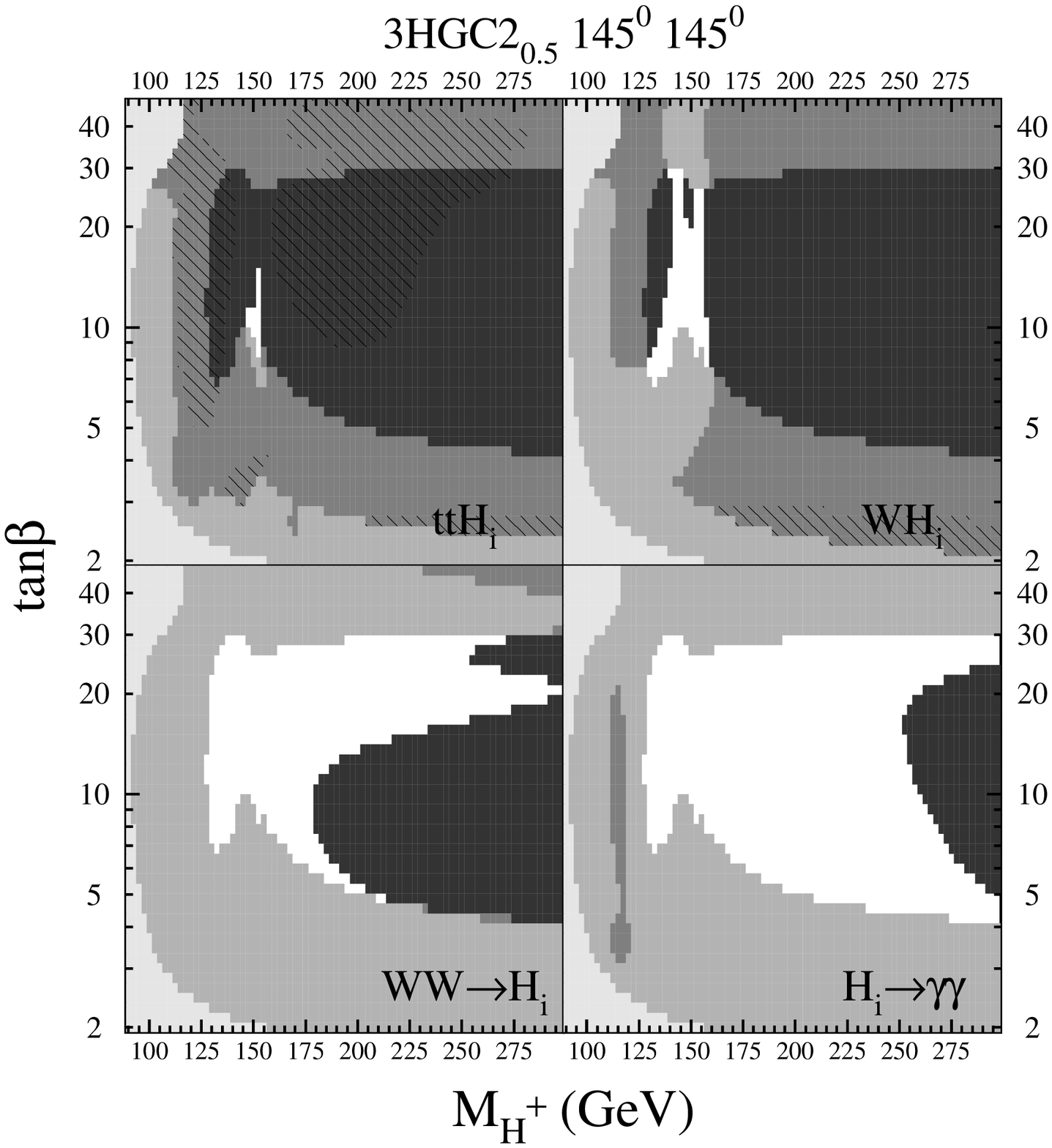}
\caption{\it Same as Fig.~\ref{com00}, but for the 3HGC2 scenario
defined in Fig.~\ref{fig:lep2}.  As in Fig.~\ref{comtry}, a region
remains uncovered due to dilution of the $g_{H_iVV}$ coupling among
three Higgs bosons.}
\label{comyyy}
  \end{center}
\end{figure}

In Figs.~\ref{com00}--\ref{comyyy}, we present a detailed analysis of
the regions covered by the relevant search channels at the Tevatron
and at the LHC. In each of these Figures, the four panels display the
regions for which Higgs bosons may be discovered at the 5-$\sigma$
level at the LHC via the $t\bar{t}H_i$ channel with 100~fb$^{-1}$
(two darkest shades of grey) 
and 30 fb$^{-1}$ ($135^\circ$ lines)
the weak-boson-fusion channel using 30~fb$^{-1}$
(two darkest shades of grey) and the gluon fusion channel with $H_i \to
\gamma\gamma$ using 100~fb$^{-1}$ (two darkest shades of grey) and 30 fb$^{-1}$
($135^\circ$ lines), as well as the region for which the Tevatron can
provide 3-$\sigma$ evidence with 5~fb$^{-1}$ (two darkest shades of grey) 
and 2 fb$^{-1}$ ($135^\circ$ lines) of integrated luminosity.  
In each case, the darker shade of grey extends beyond the LEP 
exclusion.
The
3-$\sigma$ coverage with 5~fb$^{-1}$ at the Tevatron is very similar
to the region that would be obtained for 5-$\sigma$ discovery with
15~fb$^{-1}$, or for 95\% C.L. exclusion with 2 fb$^{-1}$. The reason
why we have only analyzed 30~fb$^{-1}$ in the case of weak-boson
fusion is that a realistic analysis exists only for moderate
instantaneous luminosity, and detection in this channel might be
compromised in the different environment created by high-luminosity
LHC running. For the Tevatron, we chose to present results that can
simultaneously indicate the Higgs discovery potential under a
conservative assumption of the total integrated luminosity
(5~fb$^{-1}$), for which only a 3-$\sigma$ evidence may be obtained in
the regions not excluded by LEP.  This can be directly compared to a more optimistic
assumption (15~fb$^{-1}$), for which discovery may be possible well beyond
the region of parameters covered by LEP.
Finally, in all the scenarios studied, there are regions where
at least two Higgs bosons are close in mass and have similar
strength couplings to $W/Z$ bosons.  Our prescription is to
add the signals from $H_i\to b\bar b$ and
$H_i\to\tau^+\tau^-$ with no degradation if the mass difference is
less than 5 GeV (which is substantially smaller than the
expected mass resolution), otherwise the signals are treated separately.
We discuss this more later.
Because of the precision electromagnetic calorimetry expected
at the LHC, we do not
combine $H_i\to\gamma\gamma$ signatures, though this is of
little practical consequence.

Figure~\ref{com00} demonstrates the case of vanishing phases. Because
the CPX scenario has large values for $\mu$ and $m_{\widetilde g}$,
there is a large value for the supersymmetric loop corrections to the 
$b$-quark
Yukawa coupling from (\ref{dhb}, \ref{Dhb}).  Furthermore, since
sgn$(\mu m_{\widetilde g})>0$, the $b$ quark Yukawa coupling is decreased
relative to the $\tau$ lepton Yukawa coupling.  
The right lobe of the $t\bar tH_i$ coverage (a) arises from $H_i=H_1$,
with a transition to $H_2$ and then $H_3$ in the left lobe.
The suppression of the $b\bar b$ coupling relative to $\tau^+\tau^-$,
in conjunction with the sharing of the $g_{H_i\bar{t} t}$ coupling between
(in this case) two Higgs bosons, results in the dip near 
$M_{H^\pm}\sim 150$ GeV.  For this case without explicit
CP violation, that region corresponds exactly to $M_A\sim m^{\rm max}_h$,
where $m^{\rm max}_h$ is the maximal value of the Standard Model--like
Higgs boson in the decoupling limit.  
Because of the enhanced
$\tau^+\tau^-$ coupling relative to $b\bar b$, the $WW\to h$ channel
(c) does not exhibit this phenomena.  
For $\tan\beta >$ a few, the $W/ZH_i$ channel (b) has a similar
behavior as $t\bar tH_i$.
The limited coverage from the
$h\to\gamma\gamma$ channel (d) is a common feature of all the scenarios
we have studied.  Even when $g_{H_iWW}\sim 1$, 
a suppression of BR$(h\to\gamma\gamma)$
occurs because $g_{H_i\bar{b} b}$ is enhanced over the SM value.
The large values of $|\mu|$ and $|A|$ necessary to cause substantial
effects typically increased the off-diagonal elements of the
$3 \times 3$ Higgs squared--mass matrix relative to the diagonal elements.
Note that the $WW\to H_i(\to\tau^+\tau^-)$ channel with 30 fb$^{-1}$
is sufficient alone to cover the $M_{H^\pm}$--$\tan\beta$ plane.
With the same integrated luminosity, neither the $t\bar tH_i$
nor the $H_i\to\gamma\gamma$ channels even extend {\it beyond}
the LEP coverage.  Of course, the former is a discovery region,
whereas the latter is only 95\% C.L. exclusion.

Another possible variant on the
CPX scenario is to introduce CP violation to the Higgs sector
through a substantial phase for $m_{\widetilde g}$, $\phi_{\widetilde
g}=90^\circ$ (not shown).
The primary effect is to decrease the suppression of $g_{H_i\bar{b}b}$
relative to $g_{H_i\tau\tau}$, but, as discussed before,
the gluino phase effects are quite mild.  

The introduction of a phase for $A_t=A_b$ allows for CP-violation effects
at small values of $\tan\beta$, as shown in Fig.~\ref{com900}. As
discussed previously, it is now possible for the lightest Higgs boson $H_1$ to
become quite light and have a negligible coupling to vector bosons.  
Furthermore, the decay $H_2\to H_1H_1$ becomes kinematically
possible, introducing a
region that remains uncovered by all experiments if one considers only the
detailed analyses performed so far. 
Another region of difficult coverage
appears in the $t\bar t H_i(\to b\bar b)$ channel, for a charged--Higgs
mass close to 125 GeV.  
This did not occur for vanishing phases, and only arises now
because of the transition away from the region where $H_2\to H_1H_1$
occurs when $H_2$ has a substantial coupling to the $Z$ boson.
This does not occur for weak boson fusion in the
$H_i\to\tau\tau$ channel because of the
enhancement of the $\tau$ coupling over that of the $b$.
In this region of
parameters, the coupling of $H_1$ and $H_2$ to the top quark and weak
bosons is somewhat suppressed, though $H_2$ is relatively more strongly
coupled to these particles than $H_1$.  Due to finite quantum corrections,
the ratio of branching ratios BR$(H_{1}\to b\bar b)$/BR$(H_1 \to
\tau\tau)$ is suppressed with respect to the Standard Model value.  The
substantial increase in BR$(H_2 \to\tau\tau)$ means that
$H_{2}\to\tau\tau$ can still be observed, even though its vector--boson
coupling is suppressed. Indeed, $H_2$ is also visible in the channel $t
\bar{t} H_2$ in a small region of parameters, for somewhat larger values
of the charged Higgs mass.

We show in Fig.~\ref{com9090} the effect of including a gluino phase as
well, which, as mentioned before, does not change significantly the
picture, apart from the fact that small regions of uncovered parameter
space appear. Those uncovered regions appear for a charged--Higgs mass
around 130~GeV (160~GeV) because two of the Higgs bosons, $H_3$ and $H_2$
($H_1$ and $H_2$) are separated in mass (greater than 5 GeV apart) and share
equal couplings, of order a half of the SM values, to the weak bosons and
the top quark.

An interesting feature appears for the case displayed in 
Fig.~\ref{com13590}.  
For moderate and large values of the charged--Higgs mass and large values
of $\tan\beta \simgt 30$, the lightest Higgs--boson mass is about 123 GeV.
For these values of the Higgs mass, discovery in the weak-boson fusion
process demands a slightly larger branching ratio into $\tau$ leptons than
in the SM. This occurs for smaller values of the charged--Higgs mass, but
as the charged--Higgs mass increases the branching ratio into $\tau$
leptons becomes smaller than the SM one, and remains like this until the
decoupling limit. The $b \bar{b}$ decay branching ratio has a different
behavior, as we see in the panel showing the expected Tevatron coverage.
The mass $M_{H_1}$ is about 123 GeV in the large-$M_{H^\pm}$, large-$\tan
\beta$ limit, but BR$(H_1\to b \bar{b})$ is slightly larger than
in the SM, explaining the appearance of coverage in that
region; the lack of coverage starting near $M_{H^\pm}\sim 235$ GeV
and $\tan\beta\sim 20$ reflects the difficulty of establishing
evidence for a SM Higgs boson of mass 123 GeV. 
As usual, for larger values of the charged--Higgs mass, the
$\gamma\gamma$ mode becomes relevant as a discovery channel, since
the search for a Standard Model Higgs boson is efficient.
Note also the disappearance of an uncovered region
associated with $H_2\to H_1H_1$.

The case of phases equal to $60^\circ$, displayed in
Fig.~\ref{com6060}, is similar to the case of $90^\circ$,
although the asymptotic value of the $H_1$ mass is
somewhat lower. Most of the
features of the Higgs searches in these two cases have similar behaviors,
though occurring for different parameters,
so we do not expand on them further.

Our final example for the CPX scenario has phases
equal to $140^\circ$.  In this
case, unlike the other examples,
there is a relative suppression of the
$\tau$ branching ratio of $H_1$ in large fractions of the parameter space,
which explains the large regions of parameter space uncovered by the
weak-boson-fusion process. 
This search channel becomes efficient whenever the
Higgs mass is at the limit of the LEP reach, 
since in this case only a lower
decay branching ratio is sufficient, as seen in Table~3 of Appendix~B.
For slightly shifted values of $M_{H^\pm}$ and $\tan\beta$, instead
the $b\bar b$ coupling is suppressed relative to $\tau^+\tau^-$.
The complementarity of these channels is clearly illustrated in
panels (b) and (d).

One interesting phenomena that arises in the CPX scenario, which
is experimentally challenging, is the appearance of regions uncovered
because of $H_2\to H_1H_1$ decay.  As noted previously, there are regions
where two Higgs bosons share the coupling to $W/Z$ bosons
and $t\bar t$, but are separated in mass.  
In this `transition region' from one Higgs boson being
most SM--like to another one, the effective signal is halved.
In the MSSM without explicit CP violation, these Higgs bosons
are nearly degenerate in mass, and they are most likely
indistinguishable from two Higgs bosons with the experimental
resolution.  After including explicit CP violation, the mass
splittings can be significantly larger, and all three Higgs bosons
can share the $W/Z$ coupling.  We present here only two
example scenarios that exhibit this phenomena -- many more are
possible.
Under our prescription for adding signals occurring at masses within 5 
GeV, the last two scenarios analyzed in Figs.~\ref{comtry}
and~\ref{comyyy} show small
regions of parameters uncovered by any of the colliders. 
The size of these uncovered regions may be underestimated.
Since there are potentially 3 Higgs bosons yielding a similar
signature in these regions, and one may have to rely on experiment
to normalize the background, the Higgs signal may be entirely
washed out.  Consider, for example, the expected number
of signal and background events for the $\tau^+\tau^-$ channel
shown in Table 3.  Since $S/B \gg 1$, the accumulation
of signals in nearby bins could substantially increase the
background estimate.
Clearly, the
possibility that Higgs bosons may provide backgrounds for the other
Higgs bosons
should be understood and analyzed in more detail.

On the other hand, one consequence of sharing the coupling to $W$ and
$Z$ bosons is that all three neutral Higgs bosons may be observable at
hadron colliders in the $W/ZH_i$ and/or $WW\to H_i$ channels.  This is
an exciting possibility, since it would indicate that a 2HDM without
explicit CP violation is inadequate to describe the data.  We find no
such overlap of 5-$\sigma$ signatures from all three neutral Higgs
bosons in these channels for the scenarios presented here.
Furthermore, using the expected reach of the {\sc ATLAS} collaboration
for the $gg\to H_{\rm SM}\to ZZ^{*}\to 4\ell$ channel \cite{Atlas}, we
have checked that there is almost no coverage beyond LEP for observing
even a single Higgs boson.

\section{Conclusions}
\label{sec5} 

We have presented in this article a phenomenological analysis of 
Higgs boson searches
in Standard Model channels at the Tevatron collider and the LHC, 
for the case of the MSSM with explicit CP violation. 
We have also provided analytical expressions for
the effective couplings of
the neutral and charged Higgs bosons to fermions and to gauge bosons and 
the Higgs-boson self couplings within the MSSM
with explicit CP violation.
These expressions 
have been incorporated in the program {\tt CPHDECAY}~\cite{CPHDECAY},
used to perform our analysis.

After
considering the LEP limits and comparing our results with the published
LEP analyses, we have analyzed the reach of the Tevatron collider and the
LHC in the Standard Model search channels
for a neutral Higgs boson in the CPX scenario, as well
as in other interesting scenarios.  Our study was motivated by the fact
that the latest LEP data prove insufficient to exclude a light
MSSM Higgs boson in the CPX scenario, with a mass smaller than 60--70 GeV.
Further coverage of the MSSM parameter space with CP-violating phases
might still be possible with LEP data if searches for all the possible
decay channels of the Higgs bosons were optimized. Our analysis for the
Tevatron and the LHC is based on existing experimental simulations of the
search for neutral Higgs bosons at these colliders, and could also be
extended if more simulations were available.

We have shown that, in general, there is complementarity between the
various searches at the two
colliders for neutral Higgs bosons with Standard Model--like
properties. However, there are still
small regions of parameter space, for small values of the charged Higgs
boson mass and moderate values of $\tan\beta$, in which none of the three
neutral Higgs bosons can be detected with a high statistical
significance.  In these regions of parameters, 
one of two phenomena occurs.  First, the neutral Higgs boson with
dominant couplings to the $W$ and $Z$ bosons can
decay predominantly into channels which
contain either two neutral Higgs bosons, or a neutral Higgs boson and a
$Z$ boson.  The lighter Higgs boson has only feeble couplings to
the $W$ and $Z$ bosons and top quarks, and escapes detection both at LEP
and the hadron colliders.
Secondly, all three neutral Higgs bosons can share the coupling to
$W$ and $Z$ bosons and the top quark, resulting in three marginal
signal excesses.  The masses of the Higgs bosons may be uncomfortably close,
leading to the possibility that one ``signal'' is a background to another.
Detailed experimental simulations of these situations should
be performed in order to decide the detectability of Higgs bosons in these
regions of parameters.

Apart from these regions, the Standard Model search channels at the LHC
are adequate for discovering a neutral Higgs boson over the range of
supersymmetric parameters considered here.  The $t\bar t H_i(\to b\bar b)$
and $WW\to H_i(\to\tau^+\tau^-)$ channels are most important.  The
$gg\to H_i\to\gamma\gamma$ channel is greatly suppressed, and provides no
additional coverage.  This is a generic feature of the CPX and related
scenarios where the $b$-quark Yukawa coupling can be enhanced over its
Standard Model value.  However, the Tevatron $W/ZH_i(\to b\bar b)$ and
the LHC $WW\to H_i(\to\tau^+\tau^-)$ channels fall in a special category
for two reasons.  First, the production of the Higgs boson through a
tree--level coupling to the $W$ and $Z$ boson coupling demonstrates that
this Higgs boson contributed to the Higgs mechanism. Secondly, radiative
corrections can enhance or suppress the $b$-quark Yukawa coupling, while
not affecting the $\tau$-lepton Yukawa coupling, so that 
BR($H_i\to b\bar b$) and BR($H_i\to\tau^+\tau^-$) can be complementary
to each other.  We have presented results for 3-$\sigma$ evidence with
5 fb$^{-1}$ of luminosity at the Tevatron, and 5-$\sigma$ discovery with 
30 fb$^{-1}$ of luminosity
for the $\tau^+\tau^-$ signature at the LHC and with 100 fb$^{-1}$ for
the $b\bar b$ and $\gamma\gamma$ channels.  Our results indicate
that the $W/ZH_i(\to b\bar b)$ and $WW\to H_i(\to\tau^+\tau^-)$ channels alone
do not cover the entire range of supersymmetric parameters, even excluding
the difficult regions highlighted previously.

The restriction of $WW\to H_i(\to\tau^+\tau^-)$ results to only 30
fb$^{-1}$, which corresponds to accumulating data for a few years at a
moderate instantaneous luminosity, is motivated by the potential
deterioration of detection efficiency for this signal at a higher
instantaneous luminosity. It is notable that, when comparing the reach
with only 30 fb$^{-1}$ of data for all search channels, the $t\bar
tH_i(\to b\bar b)$ channel provides very little coverage beyond LEP -
though the LHC coverage is for discovery, whereas the LEP coverage is for
exclusion - and the $gg\to H_i(\to\gamma\gamma)$ channel provides no
discovery potential.  It is clear that the vector-boson fusion channel
should be studied more thoroughly with detailed detector simulations,
including the potential complications raised here regarding several
marginal signals in a similar kinematic region.

\newpage

\def\theequation{\Alph{section}.\arabic{equation}}
\begin{appendix}
\setcounter{equation}{0}
\section{Proper Higgs-boson self-couplings}

We exhibit here analytic expressions for the proper  Higgs-boson
self-couplings  $g^{3H}_{ijk}$,    $g^{HH^+H^-}_i$,   $g^{4H}_{ijkl}$,
$g^{2H H^+H^-}_{ij}$. The Latin indices  $i,j,k,l = 1,2,3$ attached to
the proper Higgs   self-couplings correspond to weak states  $\phi_1$,
$\phi_2$ and $a$.  

First, we list the proper trilinear self-couplings $g^{HH^+H^-}_i$ and
$g^{3H}_{ijk}$:
\begin{eqnarray}
  \label{1Hplus}
g^{HH^+H^-}_1 \!\!\!\!&=&\!\! 2s^2_\beta c_\beta\lambda_1\: +\: 
c^3_\beta\lambda_3\: -\: s^2_\beta c_\beta \lambda_4\: -\:
2s^2_\beta c_\beta\, {\rm Re} \lambda_5\: 
+\: s_\beta (s^2_\beta - 2c^2_\beta)\, {\rm Re}\lambda_6\: 
+\: s_\beta c^2_\beta {\rm Re} \lambda_7\, ,\nonumber\\
g^{HH^+H^-}_2 \!\!\!\!&=&\!\! 2s_\beta c^2_\beta \lambda_2\: +\: 
s^3_\beta\lambda_3\: -\: s_\beta c^2_\beta \lambda_4\: -\:
2s_\beta c^2_\beta \, {\rm Re} \lambda_5\: 
+\: s^2_\beta c_\beta \, {\rm Re}\lambda_6\: 
+\: c_\beta (c^2_\beta - 2s^2_\beta)\, {\rm Re} \lambda_7\, ,\nonumber\\
g^{HH^+H^-}_3 \!\!\!\!&=&\!\! 2s_\beta c_\beta\, {\rm Im}\lambda_5\: -\: 
s^2_\beta\, {\rm Im}\lambda_6\: -\: c^2_\beta\, {\rm Im}\lambda_7\,,\\[3mm]
  \label{3H}
g^{3H}_{111} \!&=&\!\! c_\beta\lambda_1\: +\:
\frac{1}{2}\,s_\beta\,{\rm Re}\lambda_6\,, \nonumber\\
g^{3H}_{112} \!&=&\!\! s_\beta\, \lambda_{34}\: 
+\: s_\beta\, {\rm Re}\lambda_5\: +\: \frac{3}{2}\,
c_\beta\, {\rm Re}\lambda_6\,,\nonumber\\
g^{3H}_{122} \!&=&\!\! c_\beta\, \lambda_{34}\: 
+\: c_\beta\, {\rm Re}\lambda_5\: +\: \frac{3}{2}\, 
s_\beta\, {\rm Re}\lambda_7\,,\nonumber\\
g^{3H}_{222} \!&=&\!\! s_\beta \lambda_2\: +\: 
\frac{1}{2}\, c_\beta\,{\rm Re}\lambda_7\,,\nonumber\\
g^{3H}_{113} \!&=&\!\! -s_\beta c_\beta\, {\rm Im}\lambda_5\: -\:
\frac{1}{2}\, (1 + 2c^2_\beta)\,{\rm Im}\lambda_6\,,\nonumber\\
g^{3H}_{123} \!&=&\!\! -2{\rm Im}\lambda_5\: -\: s_\beta c_\beta\, 
{\rm Im}\,(\lambda_6 + \lambda_7)\,,\nonumber\\
g^{3H}_{223} \!&=&\!\! -s_\beta c_\beta\,{\rm Im}\lambda_5\: 
-\: \frac{1}{2}\, (1 + 2s^2_\beta)\, {\rm Im}\lambda_7\,,\nonumber\\
g^{3H}_{133} \!&=&\!\! s^2_\beta c_\beta\lambda_1\: +\:
c^3_\beta\,\lambda_{34}\: -\: c_\beta (1 +
s^2_\beta)\, {\rm Re}\lambda_5\: +\: \frac{1}{2}\, s_\beta (s^2_\beta -
2c^2_\beta)\,{\rm Re}\lambda_6\: +\: 
\frac{1}{2} s_\beta c^2_\beta\, {\rm Re}\lambda_7\,,\nonumber\\
g^{3H}_{233} \!&=&\!\! s_\beta c^2_\beta\lambda_2\: +\:
s^3_\beta\, \lambda_{34}\: -\: s_\beta (1 +
c^2_\beta)\, {\rm Re}\lambda_5\: +\: \frac{1}{2} s^2_\beta
c_\beta\, {\rm Re}\lambda_6\: +\: \frac{1}{2}\, c_\beta (c^2_\beta -
2s^2_\beta)\,{\rm Re}\lambda_7\,,\nonumber\\
g^{3H}_{333} \!&=&\!\! s_\beta c_\beta\, {\rm Im}\lambda_5\: -\: 
\frac{1}{2}\,s^2_\beta\, {\rm Im}\lambda_6\: -\: \frac{1}{2}\,
c^2_\beta\, {\rm Im}\lambda_7\, ,
\end{eqnarray}
with $\lambda_{34} = \frac{1}{2}\, (\lambda_3 + \lambda_4)$ and
$s_\beta\, (c_\beta) = \sin\beta\ (\cos\beta)$.  Our analytic
expressions in~(\ref{1Hplus}) and~(\ref{3H}) agree well with those
presented in~\cite{CL}.\footnote{Recently, another study of Higgs
self-couplings appeared~\cite{DS}. The authors expressed their
analytic results in terms of CP-conserving mixing angles and
Higgs-boson masses, thus rendering a direct comparison with our
expressions very difficult.}

In the remainder of the Appendix,  we present new analytic results for
the  proper  quadrilinear self-couplings  $g^{4H}_{ijkl}$  and  $g^{2H
H^+H^-}_{ij}$.  These are given by
\begin{eqnarray}
  \label{4H}
g^{4H}_{1111} \!&=&\! \frac{1}{4}\, \lambda_1\,,\qquad 
g^{4H}_{1112} \ =\  \frac{1}{2}\, {\rm Re}\lambda_6\,,\qquad
g^{4H}_{1122} \ =\ \frac{1}{2}\,\lambda_{34}\: +\: \frac{1}{2}\,
{\rm Re}\lambda_5\,,\nonumber\\
g^{4H}_{1222} \!&=&\! \frac{1}{2}\, {\rm Re}\lambda_7\,,\qquad
g^{4H}_{2222} \ =\ \frac{1}{4}\,\lambda_2\,,\nonumber\\
g^{4H}_{1113} \!&=&\! -\frac{1}{2}\,c_\beta\, {\rm Im}\lambda_6\,,\qquad
g^{4H}_{1123} \ =\ -c_\beta\,{\rm Im}\lambda_5\: -\: \frac{1}{2}\,s_\beta\,
{\rm Im}\lambda_6\,,\nonumber\\
g^{4H}_{1223} \!&=&\! -s_\beta\,{\rm Im}\lambda_5\: -\: \frac{1}{2}\,c_\beta\,
{\rm Im}\lambda_7\,,\qquad g^{4H}_{2223}\ =\
-\frac{1}{2}\,s_\beta\,{\rm Im}\lambda_7\,,\nonumber\\
g^{4H}_{1133} \!&=&\! \frac{1}{2}\,s^2_\beta\, \lambda_1\: +\:
\frac{1}{2}\, c^2_\beta\, \lambda_{34}\: -\: \frac{1}{2}\, c^2_\beta\,
{\rm Re}\lambda_5\: 
-\: \frac{1}{2}\, s_\beta c_\beta\, {\rm Re}\lambda_6\,,\nonumber\\ 
g^{4H}_{1233} \!&=&\! -2s_\beta c_\beta\, {\rm Re}\lambda_5\: +\:
\frac{1}{2}\, s^2_\beta\, {\rm Re}\lambda_6\: +\: \frac{1}{2}\,
c^2_\beta\, {\rm Re}\lambda_7\,,\nonumber\\
g^{4H}_{2233} \!&=&\! \frac{1}{2}\,c^2_\beta\, \lambda_2\: +\:
\frac{1}{2}\, s^2_\beta\, \lambda_{34}\: -\: \frac{1}{2}\, s^2_\beta\,
{\rm Re}\lambda_5\: 
-\: \frac{1}{2}\, s_\beta c_\beta\, {\rm Re}\lambda_7\,,\nonumber\\
g^{4H}_{1333} \!&=&\! s_\beta c^2_\beta\, {\rm Im}\lambda_5\: -\:
\frac{1}{2}\, s^2_\beta c_\beta\, {\rm Im}\lambda_6\: -\: 
\frac{1}{2}\, c^3_\beta\, {\rm Im}\lambda_7\,,\nonumber\\
g^{4H}_{2333} \!&=&\! s^2_\beta c_\beta\, {\rm Im}\lambda_5\: -\:
\frac{1}{2}\, s^3_\beta\, {\rm Im}\lambda_6\: -\: 
\frac{1}{2}\, s_\beta c^2_\beta\, {\rm Im}\lambda_7\,,\nonumber\\
g^{4H}_{3333} \!&=&\! \frac{1}{4}\, \Gamma^{4H^+}\; ,
\end{eqnarray}
where $\Gamma^{4H^+}$ has been presented in~(\ref{4Hplus}), and 
\begin{eqnarray}
  \label{2Hplus}
g^{2H  H^+H^-}_{11} \!&=&\! s^2_\beta\, \lambda_1\: +\: 
\frac{1}{2}\, c^2_\beta \, \lambda_3\: -\: s_\beta c_\beta\, 
{\rm Re}\lambda_6\,,\nonumber\\
g^{2H  H^+H^-}_{12} \!&=&\! -\,s_\beta c_\beta\,\lambda_4\: -\:
2s_\beta c_\beta\, {\rm Re}\lambda_5\: +\: s^2_\beta\, {\rm
Re}\lambda_6\: +\: c^2_\beta\, {\rm Re}\lambda_7\,,\nonumber\\
g^{2H  H^+H^-}_{22} \!&=&\! c^2_\beta\, \lambda_2\: +\: 
\frac{1}{2}\, s^2_\beta \, \lambda_3\: -\: s_\beta c_\beta\, 
{\rm Re}\lambda_7\,,\nonumber\\
g^{2H  H^+H^-}_{13} \!&=&\! 
2 s_\beta c^2_\beta\, {\rm Im}\lambda_5\: -\: s^2_\beta c_\beta\,
{\rm Im}\lambda_6\: -\: c^3_\beta {\rm Im}\lambda_7\,,\nonumber\\
g^{2H  H^+H^-}_{23} \!&=&\! 
2 s^2_\beta c_\beta\, {\rm Im}\lambda_5\: -\: s^3_\beta\,
{\rm Im}\lambda_6\: -\: s_\beta c^2_\beta {\rm Im}\lambda_7\,,\nonumber\\
g^{2H  H^+H^-}_{33} \!&=&\! \Gamma^{4H^+}\; .
\end{eqnarray}
Notice    that  the    proper  self-couplings   $g^{HH^+H^-}_i$    and
$g^{3H}_{ijk}$ may be also  expressed in terms of $g^{2H H^+H^-}_{ij}$
and $g^{4H}_{ijkl}$, as follows:
\begin{eqnarray}
  \label{HHplus}
g^{HH^+H^-}_1 \!& =&\! 2c_\beta\, g^{2HH^+H^-}_{11}\: +\: 
s_\beta\, g^{2HH^+H^-}_{12}\,,\nonumber\\
g^{HH^+H^-}_2 \!& =&\! 2s_\beta\, g^{2HH^+H^-}_{22}\: +\: 
c_\beta\, g^{2HH^+H^-}_{12}\,,\nonumber\\
g^{HH^+H^-}_3 \!& =&\! c_\beta\, g^{2HH^+H^-}_{13}\: +\: 
s_\beta\, g^{2HH^+H^-}_{23}\,,\\
  \label{34H}
g^{3H}_{111} \!& =&\! 4c_\beta\, g^{4H}_{1111}\: +\: 
s_\beta\, g^{4H}_{1112}\,,\nonumber\\
g^{3H}_{112} \!& =&\! 3c_\beta\, g^{4H}_{1112}\: +\: 
2s_\beta\, g^{4H}_{1122}\,,\nonumber\\
g^{3H}_{122} \!& =&\! 3s_\beta\, g^{4H}_{1222}\: +\: 
2c_\beta\, g^{4H}_{1122}\,,\nonumber\\
g^{3H}_{222} \!& =&\! 4s_\beta\, g^{4H}_{2222}\: +\: 
c_\beta\, g^{4H}_{1222}\,,\nonumber\\
g^{3H}_{113} \!& =&\! 3c_\beta\, g^{4H}_{1113}\: +\: 
s_\beta\, g^{4H}_{1123}\,,\nonumber\\
g^{3H}_{123} \!& =&\! 2c_\beta\, g^{4H}_{1123}\: +\: 
2s_\beta\, g^{4H}_{1223}\,,\nonumber\\
g^{3H}_{223} \!& =&\! 3s_\beta\, g^{4H}_{2223}\: +\: 
c_\beta\, g^{4H}_{1223}\,,\nonumber\\
g^{3H}_{133} \!& =&\! 2c_\beta\, g^{4H}_{1133}\: +\: 
s_\beta\, g^{4H}_{1233}\,,\nonumber\\
g^{3H}_{233} \!& =&\! 2s_\beta\, g^{4H}_{2233}\: +\: 
c_\beta\, g^{4H}_{1233}\,,\nonumber\\
g^{3H}_{333} \!& =&\! c_\beta\, g^{4H}_{1333}\: +\: 
s_\beta\, g^{4H}_{2333}\; .
\end{eqnarray}
The above relations are useful for checking the
self-consistency  of  all the  analytic  expressions  pertinent to the
trilinear and quadrilinear Higgs self-couplings.

\end{appendix}

\newpage

\section{Experimental Simulations used in our Analysis}

We summarize here the results from previous simulations of Higgs
production at the LHC, upon which we have based our analysis of MSSM
neutral-Higgs production in the presence of CP violation.  In each Table,
we provide the minimum ratio $R$, relative to the Standard Model product
of cross section times branching ratio, for which a signal of the stated
significance could be obtained with the quoted luminosity.
We have provided more information regarding the expected number of signal
S and background B events for those cases when S/B$\sim 1$.

\begin{table*}[htbp]
  \begin{center}
    \begin{tabular}{c|ccccc}
  $M_H$ (GeV)    &    90    &  100  &   110 &    120 &   130 \\ \hline
  R$_{3\sigma}$(5 fb$^{-1}$)      &    0.63  &  0.71 &   0.88 &   1.07 & 
1.43
    \end{tabular}
    \label{tab:tev}
\caption{\it $R$ values for 3-$\sigma$ evidence for a Higgs boson in
the channel $W/ZH(\to b\bar b)$ at the Tevatron with 5~fb$^{-1}$ of
accumulated data~\cite{Carena:2000yx}.  These results are based
on combining the CDF and D\O\ data.}
  \end{center}
\end{table*}

\begin{table*}[htbp]
  \begin{center}
    \begin{tabular}{c|c|c|c|c}
$M_H$ (GeV) & S (100 fb$^{-1}$) &  B (100 fb$^{-1}$) &  
$\sigma_{\rm Gauss}$  &  R$_{5\sigma}$ (100 fb$^{-1}$) \\ \hline
100 & 147 & 223 &   9.84 &   0.51 \\
105 & 123 & 191 &   8.92 &   0.56 \\
110 & 100 & 160 &   7.91 &   0.63 \\
115 &  98 & 163 &   7.68 &   0.65 \\
120 &  96 & 167 &   7.43 &   0.67 \\
125 &  86 & 157 &   6.89 &   0.73 \\
130 &  77 & 148 &   6.33 &   0.79 
    \end{tabular}
\caption{\it The expected signal, background, Gaussian significance and 
the $R$ value for 5-$\sigma$ discovery of a Higgs boson in
the channel $t\bar tH(\to b\bar b)$ at the LHC with 100 fb$^{-1}$ of
accumulated data \cite{Drollinger:2001ym}. These results are for the
CMS experiment only. The Gaussian significance is based on a $K$ factor of 
unity for the signal~\cite{SDR}. }
\end{center}
\end{table*}

\begin{table*}[htbp]
  \begin{center}
    \begin{tabular}{c|c|c|c|c}
$M_H$ (GeV) & S (30 fb$^{-1}$) &  
B (30 fb$^{-1}$) &  $\sigma_{\rm Gauss}$  &  R$_{5\sigma}$ \\ \hline
110  &  11.1 & 3.9 & 4.1 & 1.22 \\
120  &  10.4 & 1.4 & 5.2 & 0.96 \\
130  &  8.6  & 0.9 & 5.0 & 1.00 \\
140  &  5.8  & 0.7 & 3.9 & 1.28 \\
150  &  3.0  & 0.6 & 2.3 & 2.17 \\
    \end{tabular}
\caption{\it The expected signal, background, Gaussian significance and 
the $R$ value for 5-$\sigma$ discovery of a Higgs boson in
the channel $WW\to H(\to \tau^+\tau^-)$ at the LHC with 30 fb$^{-1}$ of
accumulated data~\cite{LHC2}. A detector--level analysis
based on a fast simulation was presented in~\cite{Cavalli:2002vs}.
The results there for the combination of one 
hadronic and one leptonic decay of the $\tau^+\tau^-$
pair are slightly less significant than the original analysis, 
but they have not been combined with
the all leptonic decays.  Therefore, we have chosen to use the
original numbers, even though they are based on a parton--level analysis.
 }
  \end{center}
\label{tabletautau}
\end{table*}

\begin{table*}[htbp]
  \begin{center}
    \begin{tabular}{c|ccccccc}
$M_H$ (GeV) &      80  & 90 &  100& 110 & 120 & 130&  150 \\ \hline
R$_{5\sigma}$ (100 fb$^{-1}$) &  1.02 & 0.74 & 0.60 & 0.49 & 0.42 & 0.4 & 0.66 \\
    \end{tabular}
\caption{\it The expected signal, background,
Gaussian significance and the $R$ value for 5-$\sigma$ discovery of
a Higgs boson in
the channel $gg\to H(\to \gamma\gamma)$ at the LHC with 100 fb$^{-1}$ of
accumulated data~\cite{Lassila-Perini}. These apply to CMS only.}
  \end{center}
\end{table*}

\end{document}